\pdfoutput=1
\documentclass[twocolumn,showpacs,eqsecnum,amssymb,prb]{revtex4}
\usepackage{graphicx}

\def\cH{\hat{\cal H}}
\def\cL{{\cal L}}
\def\hsigma{\hat\sigma}
\def\bsigma{\mbox{\boldmath $\hat\sigma$}}
\def\hPsi{\hat\Psi}
\def\br{{\bf r}}
\def\bj{{\bf j}}
\def\be{{\bf e}}
\def\bv{{\bf v}}
\def\bp{{\bf p}}
\def\bA{{\bf A}}
\def\bE{{\bf E}}
\def\bR{{\bf R}}
\def\St{\mathrm{St}}
\def\diverg{\mathrm{div}}

\begin{document}

\title{Effect of radiation on transport in graphene}
\author{S.~V. Syzranov$^1$, M.~V. Fistul$^1$, and K.~B. Efetov$^{1,2}$}
\affiliation{$^1$Theoretische Physik III, Ruhr-Universit\"at
Bochum, D-44801 Bochum, Germany\\
$^2$L. D. Landau Institute for Theoretical Physics, 119334 Moscow,
Russia}
\date{\today}

\begin{abstract}
We study transport properties of graphene-based p-n junctions
irradiated by an electromagnetic field (EF). The resonant
interaction of propagating quasiparticles with external
monochromatic radiation opens \emph{dynamical gaps} in their
spectrum, resulting in a strong modification of current-voltage
characteristics of the junctions. The values of the gaps are
proportional to the amplitude of EF. We find that the transmission
of the quasiparticles in the junctions is determined by the
tunneling through the gaps, and can be fully suppressed when
applying a sufficiently large radiation power. However, EF can also
generate current, but not only suppress it.
 We demonstrate that
if the height of the potential barrier exceeds a half of the photon
energy, the directed current (\emph{photocurrent}) flows through the
junction without any dc bias voltage applied. Such a photocurrent
arises as a result of inelastic quasiparticle tunneling assisted by
one- or two-photon absorption. We calculate current-voltage
characteristics of diverse graphene-based junctions and estimate
their parameters necessary for the experimental observation of the
photocurrent and transmission suppression.
\end{abstract}

 \maketitle

\section{Introduction}

Since its first fabrication \cite{Novoselov:firstgraphene},
graphene, a monolayer of carbon atoms, has attracted a lot of
attention as a candidate for the base material in nanoelectronics.
The unique feature of this two-dimensional (2D) semiconductor is the
absence of the gap between the conduction and the valence bands
\cite{Wallace:firstgap}, which allows one to change the type of the
carriers and vary their density in a wide range by applying an
external gate voltage. Other outstanding properties of the material
are the high carrier mobility and a submicron mean free path at room
temperature \cite{Geim:review, Novoselov:firstgraphene,
Morozov:twodeghall, Novoselov:qh, Zhang:qh, Das:topgating,
Morozov:mob, Wu:fet, Bolotin:mobilsuspend}. As a result, a lot of
experimental activity has been devoted recently to investigation of
possibilities of using graphene in the field-effect-transistor type
applications \cite{Lemme:fet, Huard:stanfordexp, Williams:harvardpn,
Ozyilmaz:columbiapn, Chen:fet, Wu:fet, Gorbachev:airbridge,
Liu:air}.

In order to construct such devices one needs to be able to control
and switch off currents applying gate potentials. In conventional
semiconductors currents can be blocked by applying gate voltages
that create sufficiently high potential barriers confining
electrons in a certain region. However, the dynamics of low-energy
quasiparticles in graphene is described by the zero-mass
Dirac-type Hamiltonian \cite{Semenoff:firstdirac,
Haldane:firstdirac}, making possible their {\it reflectionless}
transmission through potential barriers of arbitrary strength (the
so-called Klein paradox) \cite{Katsnelson:klein}. Such an unusual
phenomenon reduces the possibility to switch off currents by means
of gate electrodes and is a very important property of graphene.

It is also well known for many years that the transport properties
of diverse semiconducting nanostructures are sensitive to an
externally applied electromagnetic field (EF). Though optoelectronic
devices such as radiation-controlled field-effect transistors,
photodiodes, and light-emitting diodes have been fabricated on the
basis of carbon nanotubes \cite{Duan:ligtdiode, Marcus:photofet,
Freitag:photofet, Ohno:photofet, Wang:plpc}, a little attention has
been paid so far to the possibility to control and switch off
currents in graphene using external electromagnetic radiation.

In this paper we study theoretically the transport in graphene
subject to a coordinate-dependent potential $U(\br)$ and
irradiated by electromagnetic field. For simplicity we assume that
EF is monochromatic and linearly polarized but a proper
generalization can be easily made.
 We show that the resonant
interaction of propagating quasiparticles in graphene with EF leads
to the formation of a {\it dynamical gap} $\Delta$ in the
quasiparticle spectrum. The value of the gap depends on the
intensity $S$ and the frequency $\Omega$ of external radiation. The
formation of such a dynamical gap is a generic quantum property of
systems described by the two-bands Hamiltonian, and the quantity
$\Delta/\hbar$ has the same meaning as the famous Rabi frequency for
microwave-induced quantum coherent oscillations between two energy
levels \cite{Hanggi:floquet}. For conventional semiconductors the
dynamical gap has been predicted in the
Ref.~\onlinecite{Galitski:firstgap}. Charged impurities in a sample
\cite{Elesin:imp}, electron-phonon \cite{Elesin:elphon}, and
electron-electron interactions \cite{Elesin:phonrecombee} smear the
quasiparticle spectrum and wash out the dynamical gap. Therefore,
the dynamical gap in a uniform semiconductor can be observed only if
the radiation intensity exceeds some critical value $S_c$. Under
such conditions it has been observed in the optical experiments
\cite{Elesin:gapobs, Vu:gapobs}.

So far only bulk properties (such as conductivity
\cite{Goreslavskii:cond} and coefficient of light absorption
\cite{Galitski:firstgap, Alexandrov:absorption, Inagaki:BEG}) in
absence of coordinate-dependent potentials have been studied in
semiconductors with dynamical gaps. In this case an experimental
observation of a dynamical gap is quite difficult and demands
considerable efforts. At the same time, the question about what
happens if the system is subject to an external potential has not
been addressed yet.

In this paper we consider irradiated graphene junctions subject to
a {\it non-uniform} step-like external potentials, such that the
energies of conducting electrons are located inside the dynamical
gaps only in some narrow resonant regions. This means that during
the transmission through the junction each electron in a large
energy interval of the order of potential height has to pass
through a small ``classically forbidden'' spatial region, where it
has to move inside the gap. Thus, the current-voltage
characteristics of the junctions are mainly determined by the
tunneling through the gaps.

Changing the values of the gaps (i.e. the intensity and the
frequency of external radiation) and the heights of potential
barriers, it is possible to vary the current-voltage
characteristics, e.g. to suppress Klein tunneling in graphene p-n
junctions by a sufficiently strong radiation. A short account of
this phenomenon has been considered recently in
Ref.~\onlinecite{Fistul:gap}.

However, it turns out that even more interesting effect is possible
in such a p-n junction. Here we demonstrate that in this system,
provided the height of the potential is larger than a half of the
photon energy, the directed current (\emph{photocurrent}) flows
through the junction without any dc bias voltage applied. Such a
photocurrent arises as a result of inelastic quasiparticle
transmission through the junction assisted by one- or two-photon
absorption.
 We show, that in the presence of impurities and electron-electron interaction
 in the limit of
small radiation intensities the coefficients of reflection or
transmission at the resonant regions slightly differ from those in
ballistic case even if the radiation intensity $S$ is much below
the critical value $S_c$, at which the dynamical gap vanishes. As
a result, for experimentally relevant parameters of the junctions
the photocurrent is not affected by elastic impurities and
electron-electron interactions and, we hope, it can be observed
rather easily at moderate radiation intensities.

The photocurrent we calculated for a typical graphene p-n
junctions is by several orders of magnitude larger than those
measured in carbon nanotubes \cite{Freitag:photofet,
Ohno:photofet, Marcus:photofet} for the same radiation
intensities, since the effectively two-dimensional graphene
junction has accordingly larger number of conducting channels. As
there is no gap between the conduction and the valence bands in
graphene, and the level of doping can be varied in a wide range
using gate electrodes, graphene-based photodiode can be operated
in a wide frequency range of the external radiation in the
far-infrared region.

The paper is organized as follows. In Sec.~\ref{sec:genform}, we
derive a general expression for the current flowing through the
junction in presence of a monochromatic linearly polarized
radiation. The current is determined by the symmetries of the
Hamiltonian and the scattering matrix for the system.
 In order to find explicitly the scattering matrix
in a spatially inhomogeneous potential in the presence of radiation
one needs to know the probability of tunneling through the dynamical
gap, which is calculated in Sec.~\ref{sec:tunneling}. Then, in
Sec.~\ref{sec:ballcurrent}, we consider
 electron ballistic
trajectories in a p-n junction to determine how many times and where
electrons tunnel during the transmission through the junction. Using
these results, we calculate in Sec.~\ref{sec:photocurrent}
current-voltage characteristics of irradiated ballistic graphene
junctions. In Sec.~\ref{sec:disorder}, using the kinetic equation
approach, we analyze effects of disorder and electron-electron
interaction on the photocurrent in graphene p-n junctions. In
Sec.~\ref{sec:experiment} we estimate the magnitudes of the
photocurrent and the tunneling probabilities for experimentally
achievable radiation intensities and parameters of the junctions.

\section{Current-voltage characteristics of irradiated ballistic graphene
junctions: generic analysis}

\label{sec:genform}

In this section we derive a general expression for the
current-voltage characteristics of an irradiated two-dimensional
graphene strip connected to two reservoirs, from now on called the
left (L) and the right (R) leads. We assume that electrons in the
strip move in a coordinate-dependent potential $U(z)$, varying only
along the strip, and interact with a time-periodic EF. In the
present and two subsequent sections we consider a clean system, such
that the transport in the strip is purely ballistic.

 Another important
assumption we use is that far in the leads the radiation is either
absent or its effect on the electron motion and distribution
functions is negligible. In the next section we show that the latter
condition is satisfied since the majority of the conducting
electrons in the reservoirs have their energies not too close to the
positions of the radiation-induced gaps in their spectra.

Our derivation is similar to the one presented in the
Ref.~\onlinecite{Moskalets:cool} but is adopted for the case of
two-dimensional graphene strip and leads. Assume, that far in the
$\alpha$-th lead the electron states are plane waves, incoming to
the strip, $|\varepsilon,\theta\rangle_\alpha^{in}$, or outgoing
from it, $|\varepsilon,\theta\rangle_\alpha^{out}$, characterized
only by the energy $\varepsilon$ and the angle $\theta$ of incidence
or scattering correspondingly. Note here, that there is no
intervalley and spin scattering in a ballistic sample and thus we
disregard these degrees of freedom.

According to the Floquet theorem \cite{Hanggi:floquet}, the general
solution of the Schr\"odinger equation for an electron, moving in a
static potential and subject to a periodic perturbation, takes the
form $\Psi(t)=e^{-i\varepsilon t}\Phi_T(t)$, where $\Phi_T(t)$ is a
periodic function having the same period $T$ as the perturbation.
Therefore, an electron incident with energy $\varepsilon$ on the
strip can scatter only into the states with energies
$\varepsilon+n\hbar\Omega$, where $\Omega=2\pi/T$ and $n=0, \pm1,
\pm2, \pm3, \ldots$, or, in other words, the particle can gain or
lose only an integer number of quanta $\hbar\Omega$.

Thus, the state $|\varepsilon,\theta\rangle_\alpha^{in}$ scatters
into
\begin{eqnarray}
 \sum_{\beta,n}\int d\phi\: t_{\alpha\beta}(\varepsilon,\theta;\varepsilon+n\hbar\Omega,\phi)
 |\varepsilon+n\hbar\Omega,\phi\rangle_\beta^{out},
\end{eqnarray}
where we have introduced the amplitude $t_{\alpha\beta}$ of
scattering from the state of an electron incident on the strip from
the $\alpha$-th lead at the energy $\varepsilon$ and angle $\theta$
into the state outgoing from the strip in the $\beta$-th lead at the
energy $\varepsilon+n\hbar\Omega$ and angle $\phi$. The quantity
$|t_{\alpha\beta}(\varepsilon,\theta;\varepsilon+n\hbar\Omega,\phi)|^2d\phi$
gives the probability to scatter into the angle interval
$(\phi;\phi+d\phi)$.

Accordingly,
\begin{eqnarray}\label{tone}
 \sum_{\beta,n}\int d\phi\:
 |t_{\alpha\beta}(\varepsilon,\theta;\varepsilon+n\hbar\Omega,\phi)|^2=1.
\end{eqnarray}

In order to obtain the current flowing through the strip we
introduce the distribution functions
$f^{in,out}_\alpha(\varepsilon,\theta)$ of electrons in the
incoming/outgoing states of the $\alpha$-th lead. In the incoming
states they simply coincide with the usual Fermi distribution
functions for a given temperature and voltages applied, and,
therefore, they depend only on the energies $\varepsilon$ but not
on the angles $\theta$ and $\phi$. Using the particle conservation
law the distribution functions of the outgoing states can be
rewritten as
\begin{eqnarray}\label{tconservation}
    f_{\alpha}^{out}(\varepsilon,\theta)=
    \nonumber \\=
    \sum_{\beta,n}\int d\phi\:
    |t_{\beta\alpha}(\varepsilon+n\hbar\Omega,\phi;\varepsilon,\theta)|^2f^{in}_{\beta}(\varepsilon+n\hbar\Omega).
\end{eqnarray}

The total current flowing through the strip takes the form
\begin{eqnarray}\label{current-1}
I=4W\int\frac{p\,dp\,d\theta}{(2\pi\hbar)^2}ev\cos\theta
\left(f_L^{in}(\varepsilon(p))-f_L^{out}(\varepsilon(p),\theta)\right),
\end{eqnarray}
where $W$ is the junction width, and $v$ is the velocity of
electrons. The coefficient $4$ in the last equation accounts for the
spin and valley degeneracies of the quasiparticle spectrum in
graphene. Using Eqs.~(\ref{tone}) and (\ref{tconservation}), we
rewrite Eq. (\ref{current-1}) as
\begin{eqnarray}\label{current-2}
 I=4W\int\frac{p\,dp\,d\theta}{(2\pi\hbar)^2}ev\cos\theta
 \sum_{\beta,n}
 \int d\phi\:
 \times \nonumber \\ \times \left(
 |t_{L\beta}(\varepsilon,\theta;\varepsilon+n\hbar\Omega,\phi)|^2f_L^{in}(\varepsilon)-
 \right. \nonumber\\\left.
 -|t_{\beta L}(\varepsilon+n\hbar\Omega,\phi;\varepsilon,\theta)|^2f_\beta^{in}(\varepsilon+n\hbar\Omega)
 \right).
\end{eqnarray}

Changing the variables in Eq.~(\ref{current-2}) and using the
transverse momentum conservation law,
$p^{in}(\varepsilon^{in})\sin\theta^{in}=p^{out}(\varepsilon^{out})\sin\theta^{out}$,
one can see that the terms with $\beta=L$ vanish. Indeed, they are
responsible for the backscattering and therefore cannot contribute
to the current through the junction.

If the time-periodic EF possesses the time-reversal symmetry
($t\rightarrow-t$), with respect to some moment of time $t=0$,
e.g. if the EF is sinusoidal, then the relation
\begin{eqnarray}\label{ttr}
 |t_{LR}(\varepsilon,\theta,\varepsilon+n\hbar\Omega,\phi)|^2=|t_{RL}(\varepsilon+n\hbar\Omega,\phi;\varepsilon,\theta)|^2
\end{eqnarray}
holds, and
 Eq. (\ref{current-2}) can be reduced to
\begin{eqnarray}\label{moskalets}
 I=4W\int\frac{p\,dp\,d\theta}{(2\pi\hbar)^2}ev\cos\theta
 \times \nonumber \\ \times
 \sum_{n}P_{LR}(\varepsilon,\varepsilon+n\hbar\Omega,\theta)\left(f_L^{in}(\varepsilon)-
 f_R^{in}(\varepsilon+n\hbar\Omega)
 \right),
\end{eqnarray}
where the function
\begin{eqnarray}
 P_{LR}(\varepsilon,\varepsilon+n\hbar\Omega,\theta)=\int d\phi\:
|t_{LR}(\varepsilon,\theta,\varepsilon+n\hbar\Omega,\phi)|^2,
\end{eqnarray}
is the probability for a particle in the state
$|\varepsilon,\theta\rangle^{in}_L$ in the left lead to be
scattered into the right lead into the state with the energy
$\varepsilon+n\hbar\Omega$.

This is the most general expression for the current that can be
derived using only the symmetries of the Hamiltonian of the system
and the particle conservation law. The total current $I$ contains
the elastic component (the term with $n=0$) and inelastic
components corresponding to the terms with $n \not= 0$. Let us
emphasize that the radiation not only induces the current due to
the inelastic processes but also leads to the modification of the
probability $P_{LR}(\varepsilon,\varepsilon,\theta)$ of the
elastic scattering. In the next sections we show that this
probability can be strongly suppressed by the external radiation,
which results in the vanishing of the current through the graphene
strip. Moreover, in systems where the inelastic scattering
violates symmetries in such a way that
\begin{eqnarray}
 P_{LR}(\varepsilon,\varepsilon+n\hbar\Omega,\theta)\neq
 P_{LR}(\varepsilon+n\hbar\Omega,\varepsilon,\theta),
\end{eqnarray}
the external radiation generates a photocurrent, i.e. the directed
current without any external voltage applied to the junction
\cite{Moskalets:cool}.

In a related paper, Ref.~\onlinecite{Fistul:gap}, the possibility
of the photocurrent has been overlooked since there was not
considered inelastic transmission of electrons through the
junction, corresponding to the terms with $n\neq 0$ in
Eq.~(\ref{moskalets}).

The probabilities
$P_{LR}(\varepsilon,\varepsilon+n\hbar\Omega,\theta)$ can be
calculated explicitly as soon as the Hamiltonian of the system
under consideration is given. In the next two sections we find the
probabilities
$P_{LR}(\varepsilon,\varepsilon+n\hbar\Omega,\theta)$ for the
quantum mechanical problem of electron motion in graphene in an
external potential $U(\br)$ and irradiated by a time-periodic EF.

\section{EF-induced dynamical gaps and electron dynamical tunneling}

\label{sec:tunneling}

To elucidate the main phenomena related to the influence of
radiation on the transport in graphene, we first consider the
modification of the quasiparticle spectra due to the presence of a
monochromatic electromagnetic wave, and their tunneling through
smooth potential barriers. For simplicity, we neglect in the next
two sections scattering due to impurities and electron-phonon
interaction.

The states of electrons in graphene are conveniently described by
the four-component wavefunctions, defined on two sublattices and
two valleys. Since we consider the transport in an infinite clean
sample, we may neglect intervalley- and spin scattering and study
the propagation of electrons for different valleys and spin
directions separately.

Electron motion in the time-dependent EF is described by the
non-stationary Schr\"odinger equation,
\begin{eqnarray}\label{timeschr}
    i\hbar\frac{\partial\psi}{\partial t}=\hat H\psi,
\end{eqnarray}
where $\hat H$ is the full Hamiltonian of the system.

Near the point where the bands of graphene touch each other (Dirac
point) a simplified Hamiltonian $\cH$ of a low-energy
quasiparticle moving in a slowly varying static potential $U(\br)$
and interacting with an external electromagnetic radiation can be
written for a single valley and for a certain direction of spin as
\cite{Mele:twophotcurrent}
\begin{eqnarray}\label{Ham}
    \cH=v\bsigma\left(\bp-\frac{e}{c}\bA(t)\right)+U(\br).
\end{eqnarray}
Here $\bp$ is the momentum of the quasiparticle, $v$-- the Fermi
velocity, and $\bsigma$-- the vector of the Pauli matrices in the
sublattice space ("pseudospin" space).

The electromagnetic radiation is taken into account choosing a
proper vector potential $\bA(t)$. For a linearly polarized
monochromatic electromagnetic wave it can be taken in the form
\begin{eqnarray}\label{A}
    \bA(t)=\frac{c}{\Omega}\bE\cos(\Omega t),
\end{eqnarray}
where $\bE$ is the amplitude of the electric field in the wave.

Substituting Eq.~(\ref{Ham}) into Eq.~(\ref{A}) and using
Eq.~(\ref{timeschr}) we obtain a complete description of electron
motion of graphene. In the next three subsections we study the
properties of solutions of the time-dependent Schr\"odinger equation
in graphene in presence of linearly polarized monochromatic
radiation.

\subsection{Dynamical gap in the quasiparticle spectra}

As it has been shown long ago, in the Ref.
\onlinecite{Galitski:firstgap}, resonant interaction of
quasiparticles with EF can lead to the formation of a dynamical
gap in the electron spectrum of semiconductor. The dynamical gap
occurs in the vicinity of the values of momentum $\bp$, determined
by the resonant condition $\epsilon_c(\bp)-\epsilon_v(\bp)=\hbar
\Omega$, where $\epsilon_{v}(\bp)$ and $\epsilon_{c}(\bp)$ are the
electron energies in the valence and conduction bands
respectively.

\begin{figure}
 \includegraphics[width=3in]{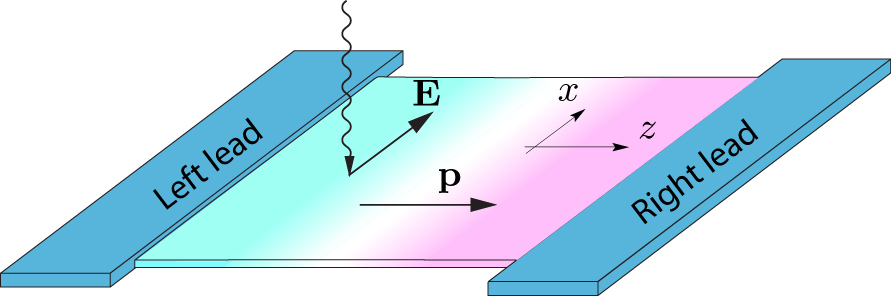}
 \caption{\label{fig:setup} (Color online)
 Graphene junction in presence of external radiation.
 Axis $z$ is directed
 along the electron momentum $\bp$ in the graphene plane, axis $x$- parallel to the electric field $\bE$ in the external electromagnetic wave.}
\end{figure}

In order to illustrate this, let us consider an electron propagating
in the absence of external potential potential
$\left[U(\br)=0\right]$ along a certain $z$ axis with momentum
$\bp$, perpendicular to the field $\bE$, which is assumed to be
directed along the $x$ axis in the plane of the graphene sheet
(Fig.~\ref{fig:setup}).

Let us emphasize, that our choice of the coordinate system and
associated pseudospin basis differs from that usually chosen in
graphene-related papers. Usually, all the calculations are done in
the basis of states, defined on different graphene sublattices, so
that the $z$ axis is directed perpendicularly to the graphene plane,
while $x$ and $y$ belong to the plane. However, for the purposes of
the present paper it is more convenient to work with the
Hamiltonian, having a diagonal form in the basis of states
\{"pseudospin parallel to the momentum $\bp$", "pseudospin
antiparallel to $\bp$"\}.

For this choice of the coordinate system the Hamiltonian
[Eq.~(\ref{Ham})] takes the form
\begin{eqnarray}\label{simpleham}
    \cH=vp_z\hat\sigma_z+2\Delta\hat\sigma_x\cos(\Omega t),
\end{eqnarray}
where
\begin{eqnarray}\label{Delta}
    \Delta=\frac{v|e|E}{2\Omega}.
\end{eqnarray}
As it will be clear from the analysis below, even a weak EF
($\Delta\ll\hbar\Omega$) strongly affects the motion of electrons
with momenta close to the resonant values,
\begin{eqnarray}\label{resmom}
    p_{res}=\pm\frac{\hbar\Omega}{2v},
\end{eqnarray}
which makes this region of momenta the most interesting. Therefore,
we consider now a quasiparticle with momentum $p_z$ close to one of
these resonant values, say, to $\hbar\Omega/(2v)$. Applying the
rotating-wave approximation (RWA) \cite{Allen:rwa}, we decompose the
second term of the Hamiltonian (\ref{simpleham}) into the right and
the left circularly-polarized waves, rotating about $z$ axis in the
pseudospin space, and neglect the second one in the vicinity of the
resonance chosen. In the other words, we make the following
replacement in Eq.~(\ref{simpleham}):
\begin{eqnarray}\label{DeltaPert}
    2\Delta\hat\sigma_x\cos(\Omega t)\longrightarrow\nonumber\\
    \Delta(\hsigma_x\cos(\Omega t)+\hsigma_y\sin(\Omega t)).
\end{eqnarray}

Then, going to the reference frame (RF) rotating counterclockwise
about the $z$ axis with the angular velocity $\Omega$ in the
pseudospin space, we come to a problem with a static Hamiltonian.
This procedure corresponds to the application of the unitary
transformation
\begin{eqnarray}\label{RF}
    \hat U=e^{i\frac{\Omega t}{2}\hsigma_z}
\end{eqnarray}
to the two-component wavefunctions. In the new basis we obtain the
time-independent effective
Hamiltonian 
\begin{eqnarray}\label{redham}
    \cH^\prime=\left(vp_z-\frac{\hbar\Omega}{2}\right)\hsigma_z+\Delta \hsigma_x.
\end{eqnarray}

The eigenvalues of the Hamiltonian (\ref{redham}) determine the
quasiparticle spectrum as
\begin{eqnarray}\label{spectra}
    \varepsilon_{e,h}(p_z)=\pm\left[\left(vp_z-\frac{\hbar\Omega}{2}\right)^2+\Delta^2\right]^\frac{1}{2}.
\end{eqnarray}

Equation (\ref{spectra}) shows that the spectrum acquires a
dynamical gap $2\Delta$ at the resonant value of momentum. This gap
is proportional to the amplitude $E$ of the radiation and inversely
proportional to its frequency $\Omega$. Far from the resonance
($|p_zv-\hbar\Omega/2|\gg\Delta$), the corresponding quasiparticles
are just conventional electrons and holes in the absence of
radiation, having spectra $\pm|vp_z-\hbar\Omega/2|$ in the chosen
rotating RF.

When the momentum $p_z$ approaches the other resonance,
$-\hbar\Omega/2v$, one can analogously calculate the quasiparticles
spectra in the RWA in the RF rotating clockwise about the $z$ axis,
\begin{eqnarray}\label{spectra2}
    \varepsilon_{e,h}(p_z)=\pm\left[\left(vp_z+\frac{\hbar\Omega}{2}\right)^2+\Delta^2\right]^\frac{1}{2}.
\end{eqnarray}

Let us emphasize that the applicability of the RWA for the
derivation of each of the Eqs.~(\ref{spectra}) and (\ref{spectra2})
is limited not by the closeness of the momentum to the corresponding
resonance, but by the negligibility of the influence of the other
resonance. For instance, Eq.~(\ref{spectra}) is valid as soon as
\begin{eqnarray}
    \left|p_z+\frac{\hbar\Omega}{2v}\right|\gg \frac{\Delta}{v}.
\end{eqnarray}

The dynamical gaps in the quasiparticle spectra have been calculated
assuming that EF was linearly polarized perpendicular to the
direction of momentum $\bp$. If the field $\bE$ is directed at some
angle $\gamma$ with respect to its transverse component $\bE_\bot$,
perpendicular to $\bp$, then the effective dynamical gap decreases
and becomes equal to
\begin{eqnarray}\label{DeltaEff}
    \widetilde\Delta(\gamma)=\Delta\cos\gamma.
\end{eqnarray}

As we will show in Sec.~\ref{sec:experiment}, for the reasonable
values of the radiation power this gap is much smaller than all the
other energy scales in the problem, such as the Fermi energy, the
photon energy $\hbar\Omega$, the heights of the potential barriers,
and the typical voltages applied to a sample.

Radiation-induced gaps have been observed in the spectra of
spontaneous radiation of conventional semiconductors subject to a
strong monochromatic laser field \cite{Elesin:gapobs, Vu:gapobs}.
In this paper we consider the effect of radiation on the {\it
transport} properties of graphene. Measurement of the current is a
simpler task and we hope that the corresponding experiments are
also possible.

In the limit of small value of $\Delta$ considered here
($\Delta\ll\hbar\Omega$), the conductivity of a homogeneous
irradiated semiconductor sample \cite{Goreslavskii:cond} or the
conductance of a tunnel junction between two irradiated uniform
samples \cite{Galitski:firstgap} can be strongly affected by the
radiation only provided the Fermi level is close to the position of
the EF-induced gap. Any considerable spatial variation of potential
due to inhomogeneities in the system smears the gap and makes its
observation in bulk semiconductors very difficult.

In the present paper we consider an essentially different situation
when electrons move in a non-uniform step-like potential and their
chemical potentials in the left and the right reservoirs are not
necessarily close to those, corresponding to the resonant momenta.
Nevertheless, for a broad interval of energies of the incident
electrons each of them achieves resonances in one or several {\it
resonant points}.

The reason for this unusual behavior is the coordinate dependence of
the electron momenta that determine the behavior of the
wavefunctions. The momenta of electrons reach their resonant values
near the step. The region, where these momenta correspond to the
motion inside the dynamical gap, is thus localized in space close to
the step-like potential. Electrons must tunnel through the gap in
order to contribute to the current between the leads.
 In the regions between the resonant
points electrons weakly interact with the radiation and propagate
freely. However, the current-voltage characteristic of such a
junction is determined by the tunneling through the dynamical gaps.

\subsection{Dynamics of electrons normally incident on irradiated potential barrier}

In this and the next subsections we study the tunneling through the
dynamical gap in the momentum space in the vicinity of the resonant
point.

\begin{figure}[h]
 \includegraphics[width=3in]{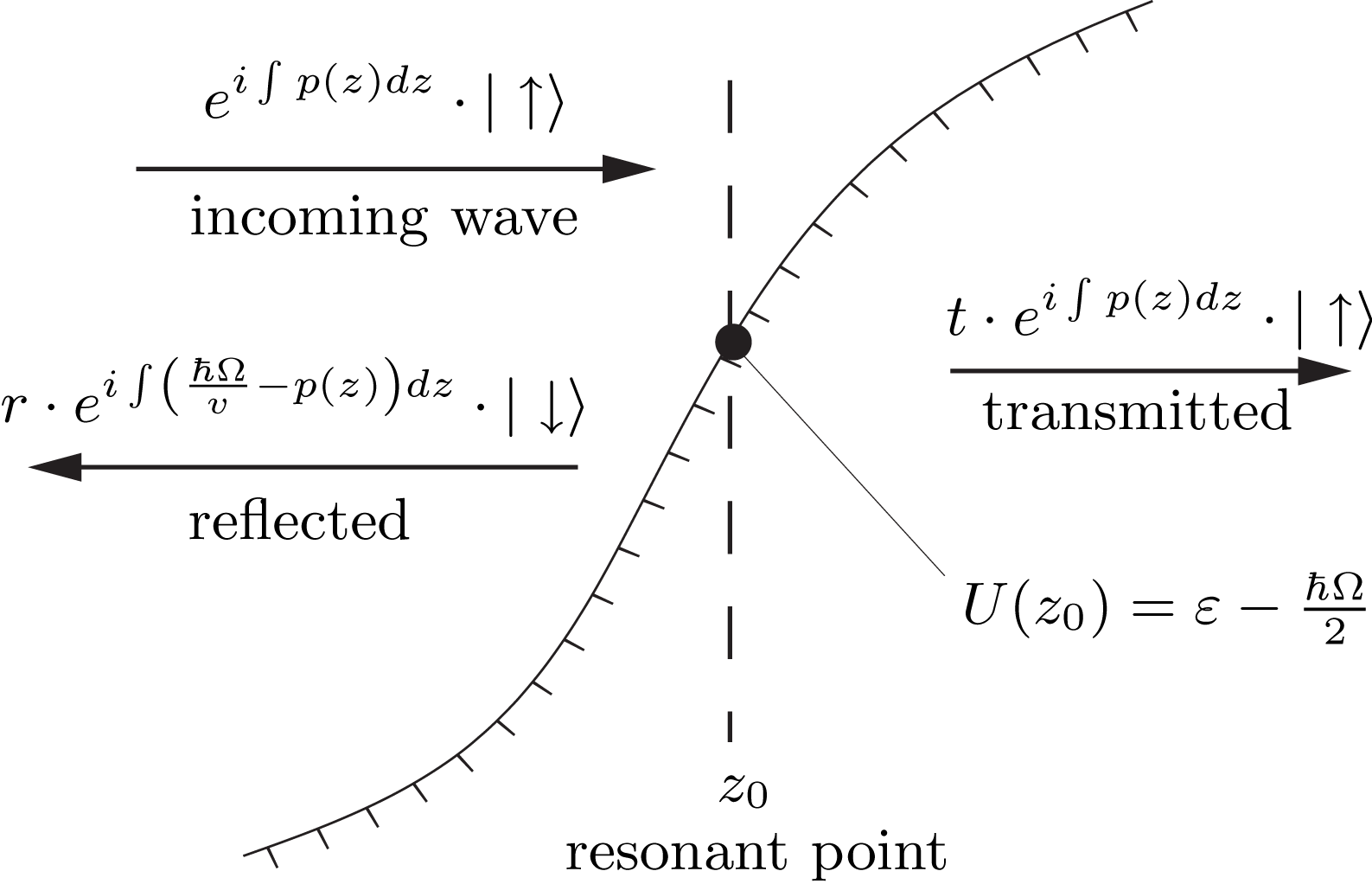}
 \caption{\label{fig:barrierone} Scattering of quasiparticles moving perpendicularly to the barrier.}
\end{figure}

 Consider first an electron
{\it normally} incident on a smooth potential barrier $U(\br)=U(z)$,
varying only in one dimension. To be specific, we assume that the
electron is moving in the conduction band along the $z$ axis, and
the height of the potential barrier increases with $z$. Let
$\varepsilon$ be the total electron energy (kinetic+potential) in
the initial laboratory RF far from the resonant point $z_0$
[determined by the equation
$U(z_0)=\varepsilon-\frac{\hbar\Omega}{2}$], where the momentum in
the absence of the radiation would equal to the resonant value
$\hbar\Omega/2v$ (see Fig.~\ref{fig:barrierone}). The EF in the
graphene plane is polarized perpendicularly to the $z$ axis, the
direction of electron momentum.

As follows from the Hamiltonian (\ref{Ham}), the electron velocity
far from the resonance, where $|U(z)-U(z_0)|\gg\Delta$, is
determined by its pseudospin:
\begin{eqnarray}
    \bv=v\bsigma.
\end{eqnarray}

This means that far from the resonant point the velocities and the
pseudospin $|\uparrow\rangle$ of the incident
$e^{i\hbar^{-1}\int^{z}p(z)dz}$ and the transmitted
$te^{i\hbar^{-1}\int^{z}p(z)dz}$ waves are both directed along the
$z$-axis. The momentum $p(z)$ entering the exponents is determined
by the equation
\begin{eqnarray}\label{pofU}
    p(z)=(\varepsilon-U(z))/v.
\end{eqnarray}
Let us emphasize, that in contrast to the corresponding
quasiclassical expression for the Schr\"odinger equation,
Eq.~(\ref{pofU}) is exact, i.e. valid for arbitrary values of
momenta $\bp$, which follows immediately from the Dirac-type
Hamiltonian (\ref{Ham}). According to our notations, the pseudospin
$|\downarrow\rangle$ of the reflected wave is antiparallel to the
$z$ axis.

 Let us find the transmission and
reflection coefficients of the electron in the region of the barrier
where it strongly interacts with the radiation. In the pseudospin
space we go to the rotating RF defined by transformation (\ref{RF}).
In this frame the Hamiltonian is static, the particle has energy
$\varepsilon-\frac{\hbar\Omega}{2}$ and scatters elastically. The
transmission and reflection coefficients are found solving the Dirac
equation for the quasiparticle wavefunctions close to the resonant
point. In the presence of the potential $U(z)$ this equation takes
the form
\begin{eqnarray}
    \left[\left(v\hat p_z-\frac{\hbar\Omega}{2}\right)\hsigma_z+\Delta\hsigma_x+U(z)\right]\Psi=\varepsilon\Psi.
\end{eqnarray}
Here $\Psi$ is the two-component wavefunction.

Without losing generality, we can set $z_0=0$. Since the potential
is smooth, it can be expanded in small $z$ and become linear around
the resonant point:
\begin{eqnarray}
    U(z)\approx\varepsilon-\frac{\hbar\Omega}{2}+Fz.
\end{eqnarray}
 The
Dirac equation for the particle with energy
$\varepsilon-\frac{\hbar\Omega}{2}$ can be written in the momentum
representation as
\begin{eqnarray}\label{schroedinger}
    -i\hbar F\frac{\partial}{\partial p}\Psi(p)=
    \left(\hsigma_z v\left(p-\frac{\hbar\Omega}{2v}\right)+\Delta\hsigma_x\right)\Psi(p).
\end{eqnarray}
We have taken into account that in the momentum representation $\hat
z=i\hbar\frac{\partial}{\partial p}$.

\begin{figure}[ht]
 \includegraphics[width=2.5in]{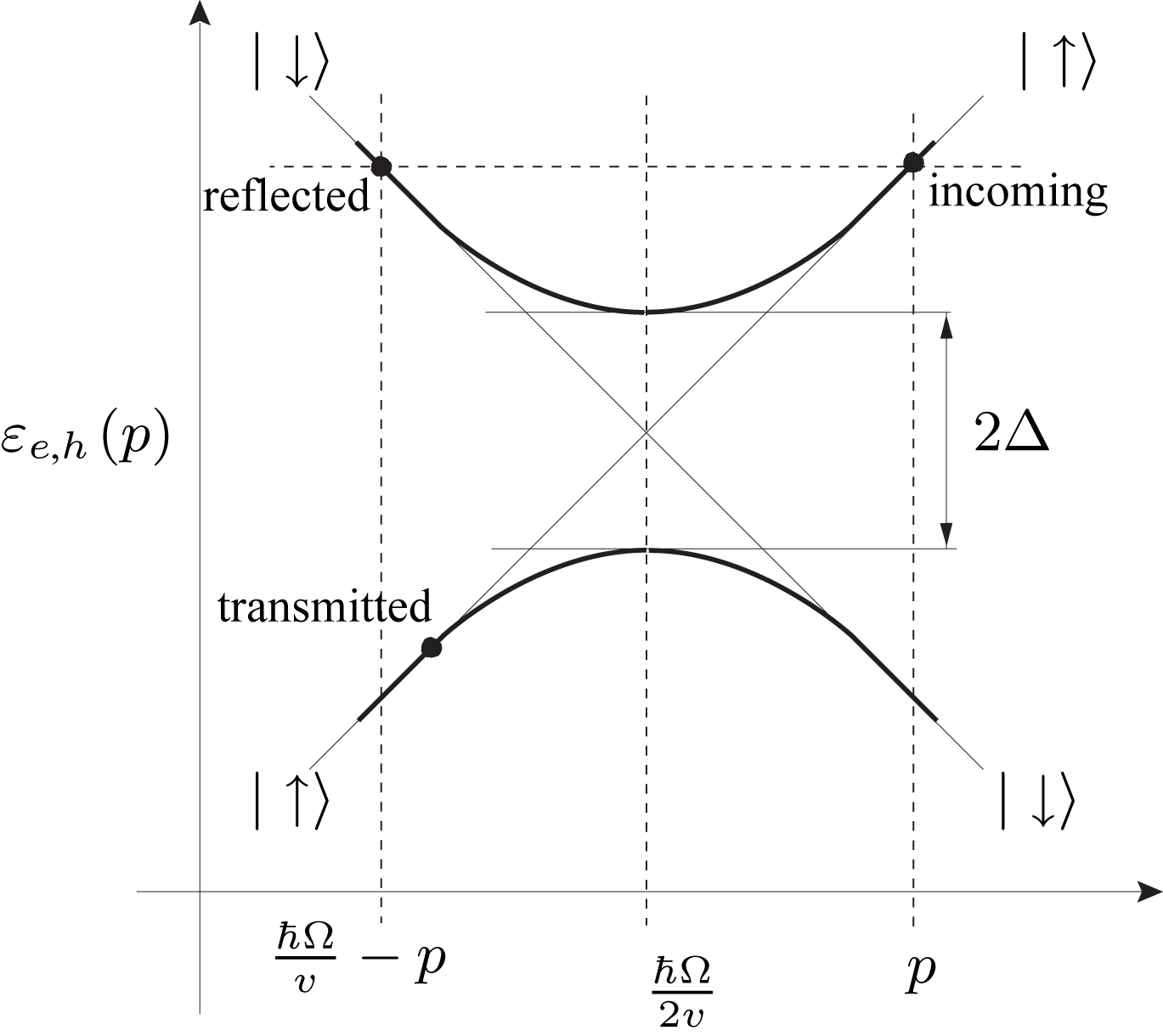}
 \caption{\label{fig:LZ} Tunneling through the dynamical gap in the momentum space (rotating RF).
 The quasiparticle spectrum in presence of radiation, Eq.~(\ref{spectra}), is shown by the solid lines in the
 plot, the spectrum in absence of radiation- by the thin lines.}
\end{figure}

Equation (\ref{schroedinger}) describes Landau-Zener tunneling
through the dynamical gap (Fig.~\ref{fig:LZ}) in the momentum space.
This phenomenon is analogous to the electron tunneling through the
forbidden band in a conventional semiconductor tunnel p-n junctions
\cite{Kane:semicondgap} or through a non-irradiated graphene p-n
junctions, when the incident electron has some finite transverse
component of momentum. The latter case, when the role of the gap is
played by the quasiparticle energy at zero longitudinal momentum,
has been studied in the Refs.~\onlinecite{CheianovFalko} and
\onlinecite{Shytov:LZmagn}.

Equation~(\ref{schroedinger}) can be solved exactly and the
tunneling probability $T=|t|^2$ takes the form
\begin{eqnarray}\label{T}
    T=e^{-\frac{\pi\Delta^2}{\hbar v F}}.
\end{eqnarray}

We emphasize that Eq.~(\ref{T}) gives the exact probability of
Landau-Zener tunneling, i.e. it is valid for arbitrary values of the
parameter
\begin{eqnarray}\label{L}
    \cL=\frac{\pi\Delta^2}{\hbar v F}
\end{eqnarray}
in the exponent rather than only in the ``quasiclassical'' limit
($\cL\gg 1$), considered in the previous papers \cite{Fistul:gap,
Fistul:rashbagap}.

According to the energy conservation law in the rotating RF, the
reflected wave has the momentum $\frac{\hbar\Omega}{v}-p(z)$ at the
same point $z$, where the incident wave had momentum $p(z)$
(Fig.~\ref{fig:LZ}). This momentum corresponds to the energy
$\varepsilon-\hbar\Omega$ of the scattered electron in the initial
laboratory RF. This means, that reflecting from the resonant point
the electron emits a photon of the energy $\hbar\Omega$.

Similarly, one can consider the cases of the quasiparticle
scattering in the valence band of graphene at the resonant points,
where the momentum in absence of radiation equals the other resonant
value $-\frac{\hbar\Omega}{2v}$.

So, we conclude that an electron normally incident on an irradiated
potential barrier can pass with the probability $T$ through the
resonant point, where its kinetic energy $vp$ in absence of
radiation would be equal $\hbar\Omega/2$ or be reflected from that
point with the probability $1-T$, emitting a photon of energy
$\hbar\Omega$. At the other type of the resonant points, where the
kinetic energy in the absence of radiation equals $-\hbar\Omega/2$,
the particle absorbs a photon of energy $\hbar\Omega$ during the
reflection. The emission/absorption is accompanied by the particle
reflection and the pseudospin-flip.

{\it Tunneling suppression.} Applying strong radiation or smooth
potential barriers, such that $\cL\gg1$, one can achieve small
tunneling probabilities $T\ll1$ [Eq.~(\ref{T})]. Such a situation
was considered in the Ref.~\onlinecite{Fistul:gap}, where it was
argued that in this case one would be able to suppress Klein
tunneling and confine electrons by irradiated barriers. In
Sec.~\ref{sec:experiment} we revisit this issue and estimate the
radiation power needed to realize such a confinement.

{\it Limit of small $\cL$}. Let us show now, that in presence of
impurities, electron-phonon and electron-electron interactions in
the sample in the limit $\cL\ll1$ the probabilities of reflection
and transmission at the resonant point remain the same as in the
ballistic sample even for the radiation intensities $S$ much below
the critical value $S_c$, at which the gap is suppressed in a wide
uniform sample. Making this statement we imply that $\cL$ is given
in Eq.~(\ref{L}) with $\Delta$ being not the real dynamical gap, but
the quantity defined by Eq.~(\ref{Delta}), which gives the value of
the gap for a clean sample neglecting impurities, electron-phonon
and electron-electron interactions.

For small $\cL$ Landau-Zener tunneling in the ballistic sample
occurs in the momentum interval of the order of
\begin{eqnarray}\label{pLZ}
    \vartriangle p_{LZ}\sim\sqrt{\frac{\hbar F}{v}}.
\end{eqnarray}
In the coordinate space the corresponding tunneling length is
\begin{eqnarray}\label{rLZ}
    \vartriangle r_{LZ}=\frac{v\vartriangle p_{LZ}}{F}\sim\sqrt{\frac{\hbar
    v}{F}}.
\end{eqnarray}
Note, that the scales (\ref{pLZ}) and (\ref{rLZ}) are determined by
the potential slope only and do not depend on the value of the gap.

Equation~(\ref{rLZ}) shows that the perturbation $\Delta\hsigma_x$
in the quasiparticles Hamiltonian (\ref{simpleham}) induces
transitions between the states $|\uparrow\rangle$ and
$|\downarrow\rangle$ in a wide spatial interval independent of
$\Delta$. This occurs even when the dynamical gap in the particle
spectrum is smeared due to the radiation-independent relaxation
processes.

The effect of the radiation-independent relaxation on the
Landau-Zener tunneling can be neglected provided the probability of
excitation due to it is much smaller than $1$, or, in other words,
the relaxation time $\tau_{R}$ is sufficiently large
\begin{eqnarray}\label{indcondition}
    \tau_R\gg\sqrt{\frac{\hbar}{vF}}.
\end{eqnarray}
This condition does not necessarily imply that the probability of
reflection $\cL=\pi\Delta^2/(\hbar vF)$ in the Landau-Zener
tunneling is much larger than the probability of an EF-independent
transition on the corresponding interval. However, using the
inequality (\ref{indcondition}) will allow us to consider the
EF-independent transitions and the tunneling independently of each
other in the calculation of the photocurrent in the subsequent
sections.

\subsection{Radiation-induced hops between trajectories at the resonant points}

In the previous subsection we considered electrons incident normally
on a potential barrier in a transverse EF. In this subsection we
analyze the scattering in the presence of radiation for arbitrary
angles between the electron momentum $\bp$ at the resonant point,
the slope of the potential $dU/d\br$, and the field $\bE$.

As it has been already mentioned [cf. Eq.~(\ref{DeltaEff})], the
effective dynamical gap close to the resonant point is
$\widetilde\Delta=\Delta\cos\gamma$, where
\begin{eqnarray}
    \gamma=\frac{\pi}{2}-(\widehat{\bp,\bE})
\end{eqnarray}
is the angle between the field $\bE$ and its transverse component
$\bE_\bot$, perpendicular to the momentum $\bp$.

 The effective potential
slope along the direction of the electron momentum at the resonant
point is $\widetilde F=|F\cos\beta|$, where
\begin{eqnarray}
    \beta=(\widehat{dU/d\br,\bp})
\end{eqnarray}
is the angle between the slope of the potential and the momentum at
the resonant point.

Then, instead of Eq.~(\ref{T}), one should use the formula
\begin{eqnarray} \label{Transm:gen}
    T=e^{-\frac{\pi\Delta^2\cos^2\gamma}{\hbar v|F\cos\beta|}}
\end{eqnarray}
for the probability for an electron to pass through the resonant
area without being scattered by the radiation. Below we will present
an explicit derivation of this result.

\begin{figure}[h]
 \includegraphics[width=3in]{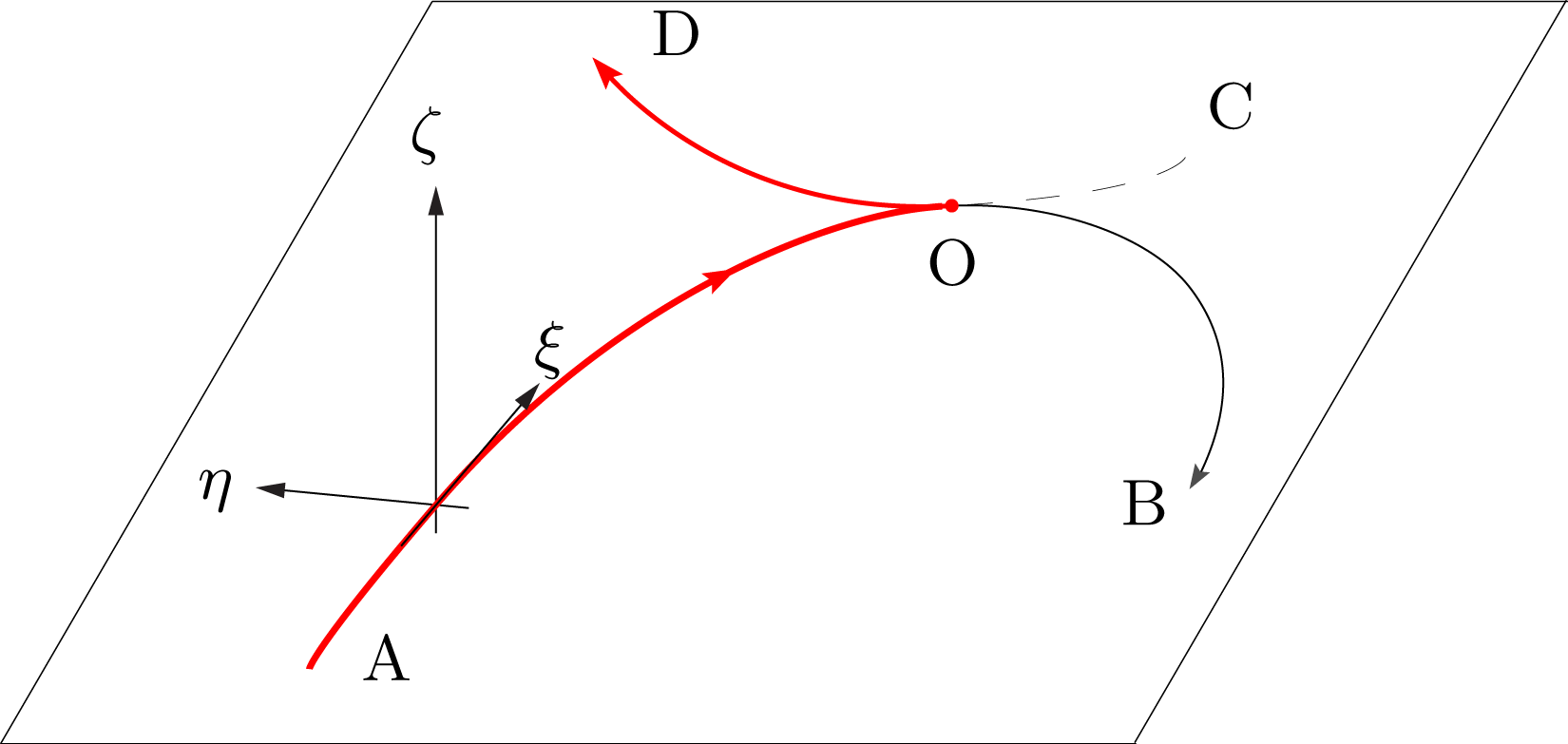}
 \caption{\label{fig:traject}(Color online) Radiation-induced hop to the trajectory with opposite pseudospin at the resonant point.
 Moving initially along the arc AO, at the resonant point O the
 quasiparticle either hops to the trajectory OD, if scattered by the radiation, or continues
 moving along OB, in case the scattering has not occurred.
 Axis $\xi$ of the coordinate system is directed along the trajectory AOB at each point, axis $\zeta$ is perpendicular
 to the graphene plane, and $\eta$ lies in the plane.}
\end{figure}

Note also that, although being reflected at the resonant point the
electron changes its pseudospin (direction of velocity), it does not
necessarily follow the same trajectory along which it moved towards
the resonant point. The reason is that the reflection is accompanied
by a photon emission or absorption, and the new electron energy
$\varepsilon_{refl}=\varepsilon\pm\hbar\Omega$ corresponds to a new
trajectory. This means that the radiation can enforce the electron
to hop onto the trajectory, corresponding to the opposite pseudospin
at the resonant point (Fig.~\ref{fig:traject}). Let us now proceed
to an explicit derivation of this result and of
Eq.~(\ref{Transm:gen}).

 As discussed in the previous subsection,
far from the resonances the electron motion can be considered
quasiclassically neglecting the radiation. Let $AOB$
(Fig.~\ref{fig:traject}) be a classical trajectory of a
quasiparticle moving in absence of radiation in the potential
$U(\br)$. Here, $O$ is the resonant point where the momentum equals
the resonant value $p=\hbar \Omega/(2v)$. At the resonant point the
particle can be reflected and hop to the other trajectory $COD$.
 Trajectory $COD$ corresponds to the same momentum but the
\emph{opposite} pseudospin at the resonant point
(Fig.~\ref{fig:traject}).

To obtain the Eq.~(\ref{Transm:gen}) and analyze the
radiation-induced hops between trajectories we
introduce at each point of AOB the coordinate system $\eta\zeta\xi$,
where the axis $\xi$ is directed tangentially along the trajectory,
$\zeta$- normally to the graphene sheet, and $\eta$ lies in the
plane of the sheet. Let $s$ be the particle coordinate along the
trajectory $AOB$. Introduce also the fixed coordinate system $xyz$,
such that the $x$ axis is directed along the EF and $y$- normally to
the graphene plane and along $\zeta$. The system $\eta\zeta\xi$ is
obtained rotating $xyz$-system around the $y$ axis by some angle
$\alpha(s)$.

The corresponding pseudospin transformation is
\begin{eqnarray}
 \hat W=e^{\frac{i}{2}\alpha(s)(\be_\zeta\bsigma)}.
\end{eqnarray}

Let us set $s=0$ at the resonant point. The Hamitonian (\ref{Ham}),
written in the pseudospin basis defined by the system $\eta\zeta\xi$
and linearized in the vicinity of the point O, takes the form
\begin{eqnarray}\label{gaugeHam}
 \tilde{H}=i\hat{\dot W}\hat W^\dagger+\hat W\cH\hat W^\dagger
 \nonumber \\
 =\hsigma_z v\left(p_\xi-\hbar\frac{\partial\alpha}{\partial
 s}\hsigma_y\right) \nonumber
 +\frac{\partial U}{\partial s}s
 \\+
 \hsigma_x vp_\eta+\frac{\partial U}{\partial\eta}\eta
 \nonumber\\
 +2\Delta(\hsigma_z\sin\alpha+\hsigma_x\cos\alpha)\cos(\Omega
 t).
\end{eqnarray}

The third line of Eq.~(\ref{gaugeHam}) describes the motion of the
wavepacket in the direction of axis $\eta$, perpendicular to the
trajectory.
This part of the Hamiltonian is independent of the longitudinal
coordinates on the scales being considered and can be disregarded if
we want to find the probability of tunneling through the point O
along the trajectory AOB.

The term $-\hbar(\partial\alpha/\partial s)\hsigma_y$ in the second
line of Eq.~(\ref{gaugeHam}) is the gauge potential generated by the
local rotations of the reference frame $\eta\zeta\xi$. Considering
the tunneling in the vicinity of the point O, one can neglect this
term, provided that the potential is smooth enough
($\lambda/v\gg\Omega$, where $\lambda$ is the characteristic scale
of the potential variation). Note, that such a neglect in general
can be made only on a small enough interval, but not on the whole
trajectory between different resonant points.

In the related paper, Ref.~\onlinecite{Fistul:gap}, the local
reference frame reverted the direction of two axes at the point
where $p=0$,
but the corresponding non-negligible delta-function-type gauge
potential, generated by this rotation, was not taken into account.

The last line of Eq.~(\ref{gaugeHam}) is the EF-induced
time-dependent perturbation of the Hamiltonian. Following the same
line of reasoning as when deriving RWA \cite{Allen:rwa}, one can
keep in this perturbation only the transverse circularly polarized
wave rotating clockwise about $\bp$ (axis $\xi$).

Making all the aforementioned approximations in the
Eq.~(\ref{gaugeHam}) and going to the rotating RF, we arrive at the
same problem as considered in the previous subsection with the
effective gap $\Delta\cos\gamma$ and the effective potential slope
$F\cos\beta$, and finally obtain the Eq.~(\ref{Transm:gen}).


\section{Dynamics of electrons in a ballistic graphene p-n junction}

\label{sec:ballcurrent}

From now on we restrict ourselves to the consideration of a specific
type of potential barriers. We assume that the potential $U(z)$,
varying only along a certain axis $z$, increases monotonically from
$U(-\infty)=0$ far in the left lead to $U(+\infty)=U_0>0$ far in the
right one. The electric field $\bE$ of the external radiation is
directed along the $x$ axis, perpendicular to the $z$ axis and lying
in the graphene plain.

From the previous section we know that the electron motion in a
non-uniform potential in the presence of radiation can be considered
as quasiclassical between the resonant points, where the hops
between trajectories can take place, accompanied by the
pseudospin-flips and photon emissions or absorptions.

Now we have to consider all the classical electron trajectories
 in the potential under consideration and
find their resonant points.

\subsection{Dynamics of electrons non-interacting with radiation}

 In the absence of radiation electron transmission through the
junction is determined by its transverse momentum
$p_\bot=p\sin\theta$, which is conserved during the motion. The
electron tunnels through the non-irradiated p-n junction with the
probability \cite{CheianovFalko}
\begin{eqnarray}
    T_0=e^{-\frac{\pi p_\bot^2v}{\hbar F_0}},
\end{eqnarray}
and, accordingly, reflects back with the probability $1-T_0$. Here
$F_0$ is the effective potential slope at the p-n interface,
calculated taking into account the charge density distribution in
the junction \cite{ZhangFogler}. For each incident electron $F_0$
should be understood as the slope of the potential at the point
where the longitudinal momentum equals zero. Note, that in general
it is different from the potential slope $F$ at the resonant point.

Let us classify the regimes of the particle motion in the absence of
radiation according to their transverse momenta. We introduce the
\emph{``2D-modes''}, with $|p_\bot|\gg(\hbar F_0/(\pi
v))^\frac{1}{2}$, and the \emph{``normal modes''}, with
$|p_\bot|\ll(\hbar F_0/(\pi v))^\frac{1}{2}$. In the first case, for
large enough $p_\bot$, the electrons perfectly reflect from the
interface, from the place where their longitudinal momentum turns to
zero. In the second case particles freely penetrate through the
junction without reflection.

The classical path of an electron in the presence of the radiation
consists of the pieces of its paths in the absence of the radiation,
stuck to each other at the resonant points.

Let us show now that on each electron trajectory in the presence of
the radiation there can be {\it no more than two resonant points}.
An electron encounters a resonant point, when its full momentum
$(p_z^2+p_\bot^2)^\frac{1}{2}$ reaches the resonant value
$\hbar\Omega/(2v)$, i.e. when the longitudinal momentum $p_z$
becomes equal to one of the two values
\begin{eqnarray}\label{reslong}
    p_{res}=\pm\left[\left(\frac{\hbar\Omega}{2v}\right)^2-p_\bot^2\right]^\frac{1}{2}.
\end{eqnarray}
The momentum $p_z$ of a classical electron, moving in the
monotonically increasing potential $U(z)$, changes in time
monotonically, $\dot{p}_z=-\partial_zU<0$. If the particle scatters
at a resonant point, its momentum does not change. Then each of the
momentum values (\ref{reslong}) of $p_z$ is reached not more than
once, and the electron encounters correspondingly not more than two
resonant points.

 In order to calculate the total current
through the junction we have to consider the electron paths starting
from the left lead and terminating in the right lead. The
contribution to the current of the inverted processes, transmission
from right to left, has been already taken into account during the
derivation of Eq.~(\ref{moskalets}).

\subsection{Electron paths in effectively two-dimensional modes.}

Let us find first the resonant points for an electron in the
``2D-mode'', i.e. having large $|p_\bot|$, and incident on the
barrier from the left with a positive energy and momentum
$p>\hbar\Omega/(2v)$.

If $|p_\bot|>\hbar\Omega/(2v)$, then there are no resonant points on
the trajectory. The electron weakly interacts with EF, perfectly
rebounds from the p-n interface in presence of radiation and thus
does not contribute to the current through the junction.

If $|p_\bot|<\hbar\Omega/(2v)$ and the potential $U_0$ is high
enough, then there are two resonant points, corresponding to the
values (\ref{reslong}) of momentum $p_z$.

The behavior of the particle is best illustrated as its path on the
plot of its kinetic energy
 in the
laboratory RF,
\begin{eqnarray}\label{kinenergy}
 \varepsilon_{kin}(p_z)=\pm
 v(p_\bot^2+p_z^2)^\frac{1}{2},
\end{eqnarray}
 versus the longitudinal momentum
$p_z$.

\begin{figure}[h]
 \includegraphics[width=1.2in]{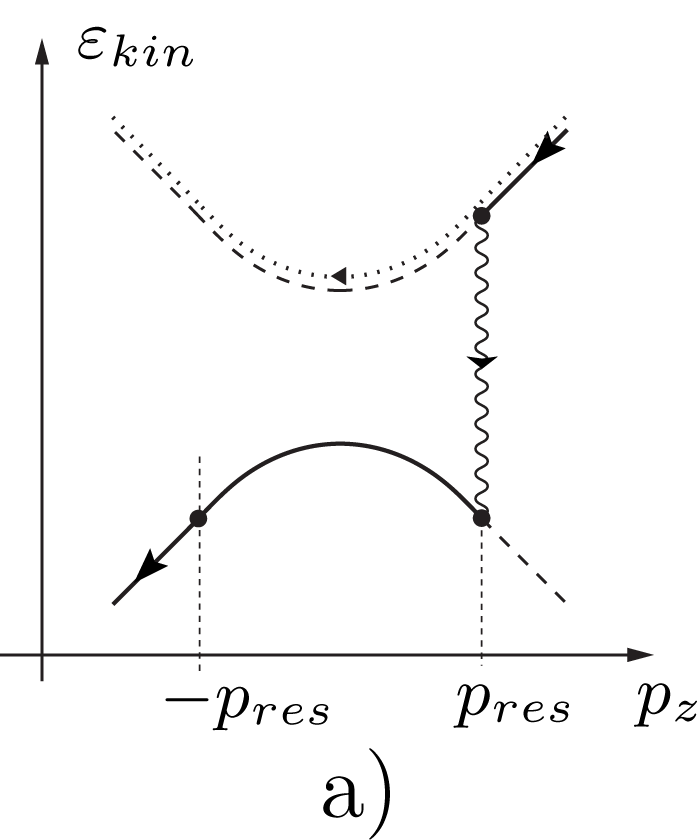}\quad\quad
 \includegraphics[width=0.5\columnwidth]{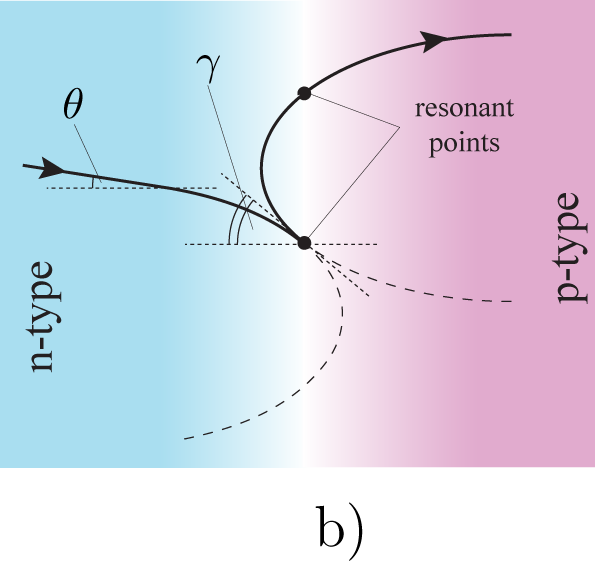}
 \caption{\label{fig:proc1} (Color online)
 Process contributing to the current through the
 junction (solid lines). The particle emits a photon at the first resonant point,
 and passes without reflection the second (``2D-modes'').
 (a) In the plot ``kinetic energy vs. longitudinal
 momentum''.
 (b) In the spatial coordinate space (graphene plane view). The emission of the photon is
 accompanied by the hop at the resonant point between two classical
 paths, obtained in absence of radiation.
 }
\end{figure}

Consider, for instance, the process illustrated in the
Fig.~\ref{fig:proc1}(a). Spectrum (\ref{kinenergy}) is shown by the
dashed lines there. The electron path on this plot, indicated by
solid lines with arrows, starts at the branch of the spectrum
corresponding to the conduction band and positive velocity
$v_z=\partial\varepsilon/\partial p_z$. If there was no radiation,
the electron would always be in the conduction band and would follow
the dotted line in the figure. In the end of the dotted line
$v_z<0$, indicating the fact that the electron would have reflected
from the junction. In fact, when reaching the first resonant point,
in the process under consideration the electron emits a photon
[curved line in the Fig.~\ref{fig:proc1}(a)], and then moves in the
valence band, passing through the second resonant point. In the end
of the process the sign of the longitudinal velocity coincides with
the initial one. This means, that as a result of the process, the
electron penetrated through the junction from one lead into another.
Thus, the radiation can assist electron transmission through the
junction.

Let us now take a look at the corresponding electron trajectory in
the coordinate space, Fig.~\ref{fig:proc1}(b). Shown there two
classical trajectories in absence of radiation are tangent at the
resonant point. The left trajectory in the absence of radiation
starts and ends up in the region of n-type graphene. It is the path
of the electron in the conduction band, having classical Hamiltonian
\begin{eqnarray}
    \cH_{cond}(\bp,\br)=vp+U(\br).
\end{eqnarray}
Similarly, the right path begins and terminates in the p-type
graphene, and corresponds to the motion of a particle in the valence
band with the Hamiltonian
\begin{eqnarray}
    \cH_{val}(\bp,\br)=-vp+U(\br).
\end{eqnarray}
Due to the effect of the radiation the particle hops at the resonant
point from the left path to the right one, the whole resulting
trajectory is shown by the solid line in the
Fig.~\ref{fig:proc1}(b).

For a given electron energy and the angle of incidence the
probability to pass along the trajectory under consideration is
\begin{eqnarray}
    P_{1scatt, 2pass}=(1-T_1)T_2,
\end{eqnarray}
where $T_1$ and $T_2$ are the transmission probabilities
correspondingly at the first and the second resonant points.

\begin{figure}[ht]
 \includegraphics[width=1.2in]{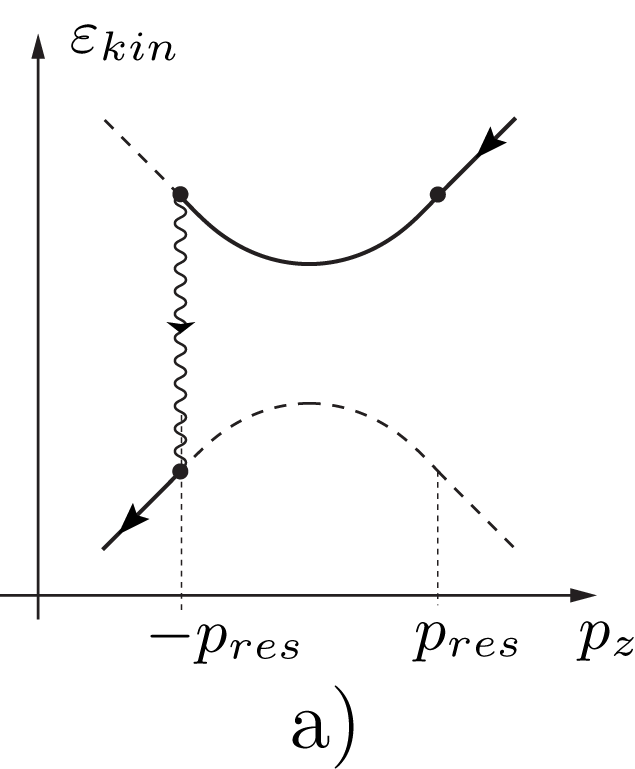}\quad\quad
 \vspace{0.1in}
 \includegraphics[width=0.5\columnwidth]{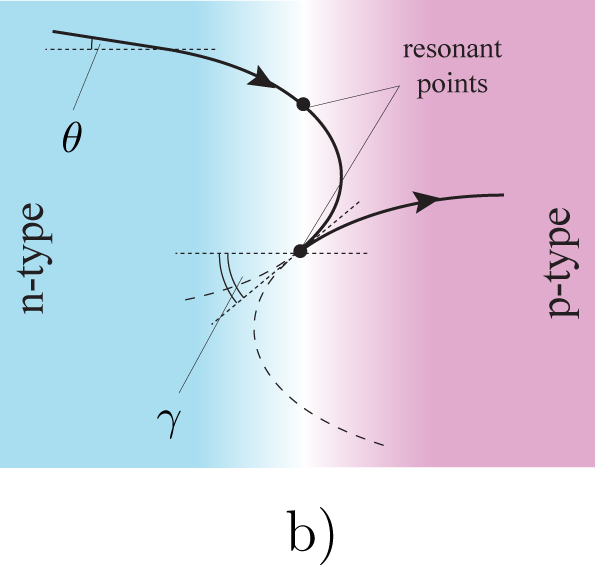}
 \caption{\label{fig:proc2} (Color online)
 Another process contributing to the current through the
 junction (solid lines).
 The particle passes without reflection the first resonant point
 and emits a photon at the second (``2D-modes'').
 (a) In the plot ``kinetic energy vs. longitudinal
 momentum''.
 (b) In the spatial coordinate space (graphene plane view). The emission of the photon is
 accompanied by the hop at the resonant point between two classical
 paths, obtained in absence of radiation.
 }
\end{figure}

 Analogous to Fig.~\ref{fig:proc1}, Fig.~\ref{fig:proc2} illustrates the
tunneling, when the particle is not scattered at the first resonant
point, but rebounds from the second one, emitting a photon.

Let us emphasize, that one should not confound the static gap
$2v|p_\bot|$ of the Hamiltonian
\begin{eqnarray}
    \cH_{kin}(p_z)=vp_z\hsigma_z+vp_\bot\hsigma_x,
\end{eqnarray}
giving the particle spectrum in the absence of radiation in the
Figs.~\ref{fig:proc1} and \ref{fig:proc2}, with the dynamical gap
$2\Delta$ in Fig.~\ref{fig:LZ}, where the quasiparticles spectrum is
shown in the rotating reference frame, taking into account the
radiation. Landau-Zener tunneling through the static gap is strongly
suppressed, since $|p_\bot|\gg(\hbar F_0/(\pi v))^\frac{1}{2}$.

\begin{figure}[ht]
 \includegraphics[width=1.5in]{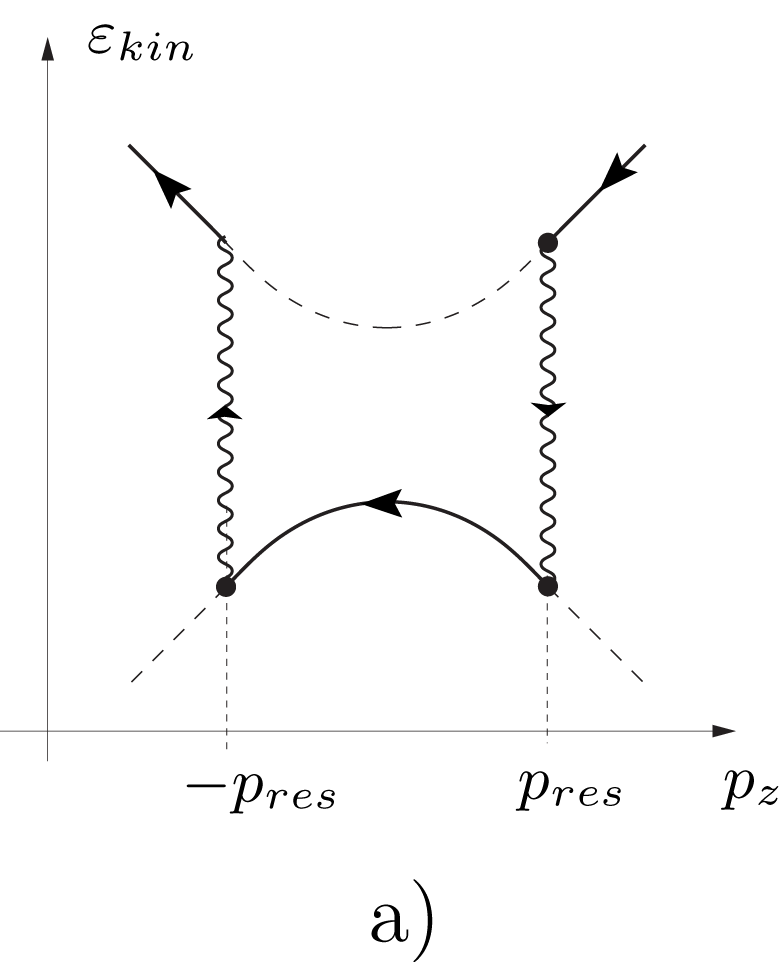}\quad\quad
 \includegraphics[width=1.5in]{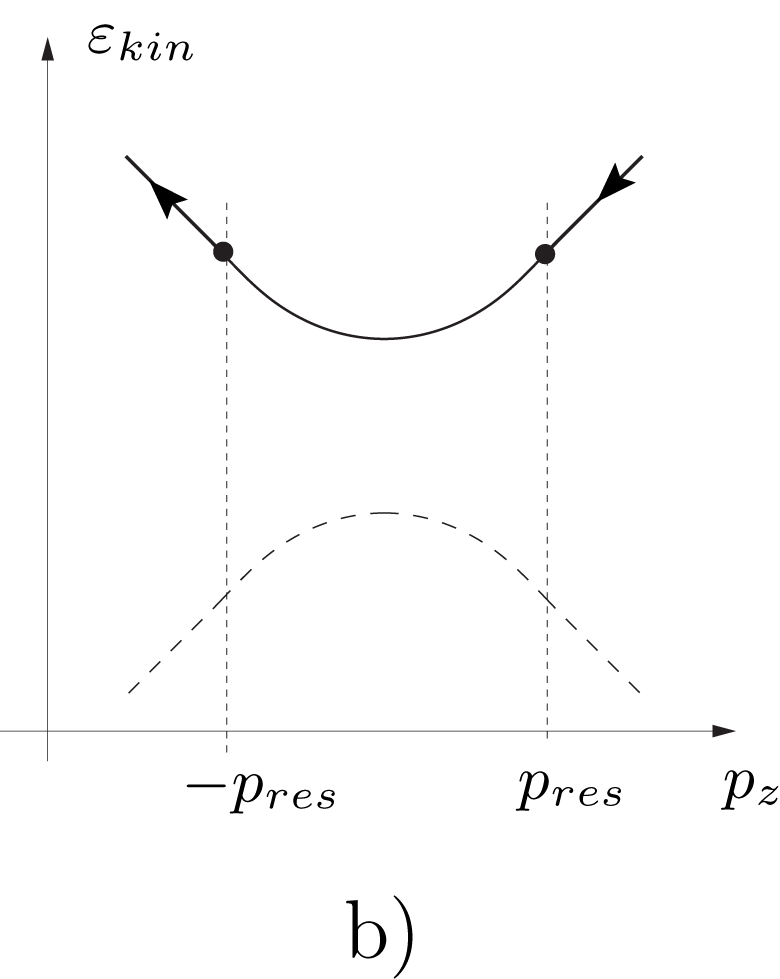}
 \caption{\label{fig:resgapLL} Kinetic energies of the particles incident from the left reservoir and
 scattered back there (``2D-modes''). a)The particle emits and absorbs a photon at the first and the
 second resonant points correspondingly. b)The particle is not scattered by the radiation.}
\end{figure}

In the processes shown in the plots in Figs.~\ref{fig:proc1}(a) and
\ref{fig:proc2}(a), electron ends up in the right lead, because
finally the particle is on the branch of the spectrum corresponding
to the same direction of velocity at $|p_z|\rightarrow\infty$ as on
the initial one. In contrast to that, Fig.~\ref{fig:resgapLL}
illustrates the backscattering: the particles enter and leave the
junction in the same lead L. According to the results of
Sec.~\ref{sec:genform}, the corresponding trajectories do not
contribute to the current and should be excluded from our
consideration.

We see that, although in the absence of radiation electrons in the
normal modes could not tunnel through the junction, they can
penetrate from the left to the right lead emitting one photon when
the sample is irradiated by EF. Due to the time-reversal symmetry,
the inverse processes, when a particle penetrates from the right to
the left lead absorbing a photon, also exist. As we discussed
before, in order to derive the current-voltage characteristics of
the junction, we have to calculate the tunneling probabilities only
for the particles incident from the left, the contribution of
inversed processes has been taken into account when deriving
Eq.~(\ref{moskalets}).

\subsection{Electron paths in normal modes.}

Now let us consider the normal modes, i.e. having such small
transverse momenta that in the momentum space the particles freely
penetrate through the static gap, as if it was zero. This
corresponds to $p_\bot=0$ in Eq.~(\ref{kinenergy}).

\begin{figure}[h]
 \includegraphics[width=1.5in]{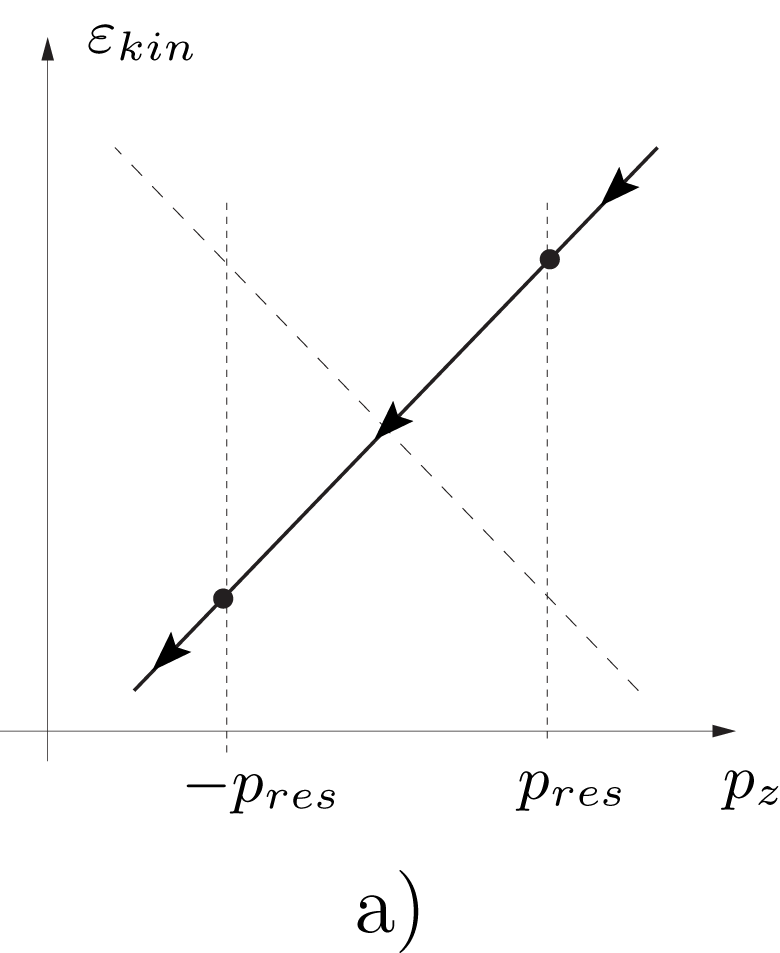}\quad\quad
 \includegraphics[width=1.5in]{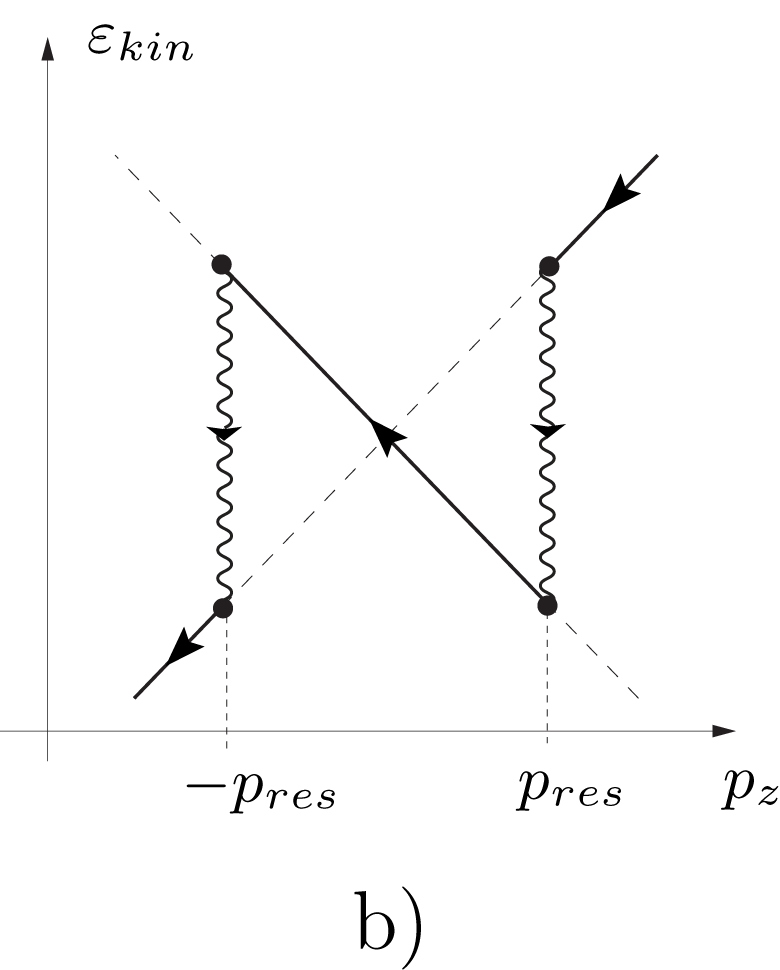}
 \caption{\label{fig:resnormal} Kinetic energies of the particles incident from the left reservoir and
 penetrating into the right one (normal modes). a)Usual Klein tunneling, no radiation-induced scattering.
 b)Tunneling accompanied by the two-photon emission.}
\end{figure}

For the tunneling from the left to the right lead both possible
electron paths $\varepsilon_{kin}(p_z)$ are shown in
Fig.~\ref{fig:resnormal}.

The first one, in the Fig.~\ref{fig:resnormal}(a), corresponds to
the usual Klein tunneling, when the particle is not scattered by the
radiation. The static gap is negligible, so the particle passes from
the conduction to the valence band through the Dirac point without
reflection.

The second path, in the Fig.~\ref{fig:resnormal}(b), illustrates the
process of tunneling from the left to the right lead, accompanied by
two photon emissions. Note, that since the particle is being twice
reflected by the EF in this process, the radiation, whatever strong,
cannot suppress this channel of tunneling. The increase in the
radiation power leads only to the enhancement of the transmission
probability in the channel.

\section{Photocurrent in ballistic samples}

\label{sec:photocurrent}

\subsection{Photocurrent due to effectively two-dimensional modes}

For the experimentally relevant parameters (see Sec.
\ref{sec:experiment} for more details), the electron Fermi momentum
in the leads is much larger than the characteristic transverse
momentum needed for a particle to be reflected at the p-n interface,
\begin{eqnarray}\label{largepf}
 p_F\gg\sqrt{\frac{\hbar F_0}{\pi v}}.
\end{eqnarray}
Hence, the majority of the particles are in the ``2D-modes''. In
this subsection we calculate the photocurrent in a ballistic
graphene p-n junction, neglecting the normal modes.

 As we
have just shown for such modes, the tunneling from left to right in
a monotonically increasing potential is necessarily accompanied by
the emission of one photon. Since the tunneling from left to right
involves photon emission, tunneling from right to left, due to the
time-reversal symmetry, involves photon absorption.

Then for some energies
\begin{eqnarray}
 P_{LR}(\varepsilon,\varepsilon-\hbar\Omega,\theta)\neq
 P_{LR}(\varepsilon-\hbar\Omega,\varepsilon,\theta)=0,
\end{eqnarray}
and, according to the results of Section \ref{sec:genform}, the
photocurrent flows through the junction in absence of any voltage
applied to it.

 If the height of the potential obeys the inequality
$U_0>\hbar\Omega/2$, then at least one of the processes in the
Fig.~\ref{fig:proc1} or Fig.~\ref{fig:proc2} or time-reversed
processes is possible in the p-n junction, and some photocurrent
flows through it. From now on we assume, however, that the barrier
height and the Fermi energy are large enough,
\begin{eqnarray}\label{phmaxcond1}
 \varepsilon_F>\hbar\Omega,\\
 U_0-\varepsilon_F>\hbar\Omega.
                \label{phmaxcond2}
\end{eqnarray}
As it will be clear from the further consideration, under these
conditions the photocurrent is maximal.

Since we are neglecting normal modes, and in the ``2D-modes'' the
tunneling from left to right is possible only with a single-photon
absorption, in Eq.~(\ref{moskalets}) we have to discard all the
terms except those with $n=-1$. Then, according to
Eq.~(\ref{moskalets}), only the electrons with energies
$\varepsilon$ in the left lead, such that
$\varepsilon-\hbar\Omega<\varepsilon_F<\varepsilon$, or, the same,
$\varepsilon_F<\varepsilon<\varepsilon_F+\hbar\Omega$, contribute to
the current.

\begin{figure}[ht]
 \includegraphics[width=\columnwidth]{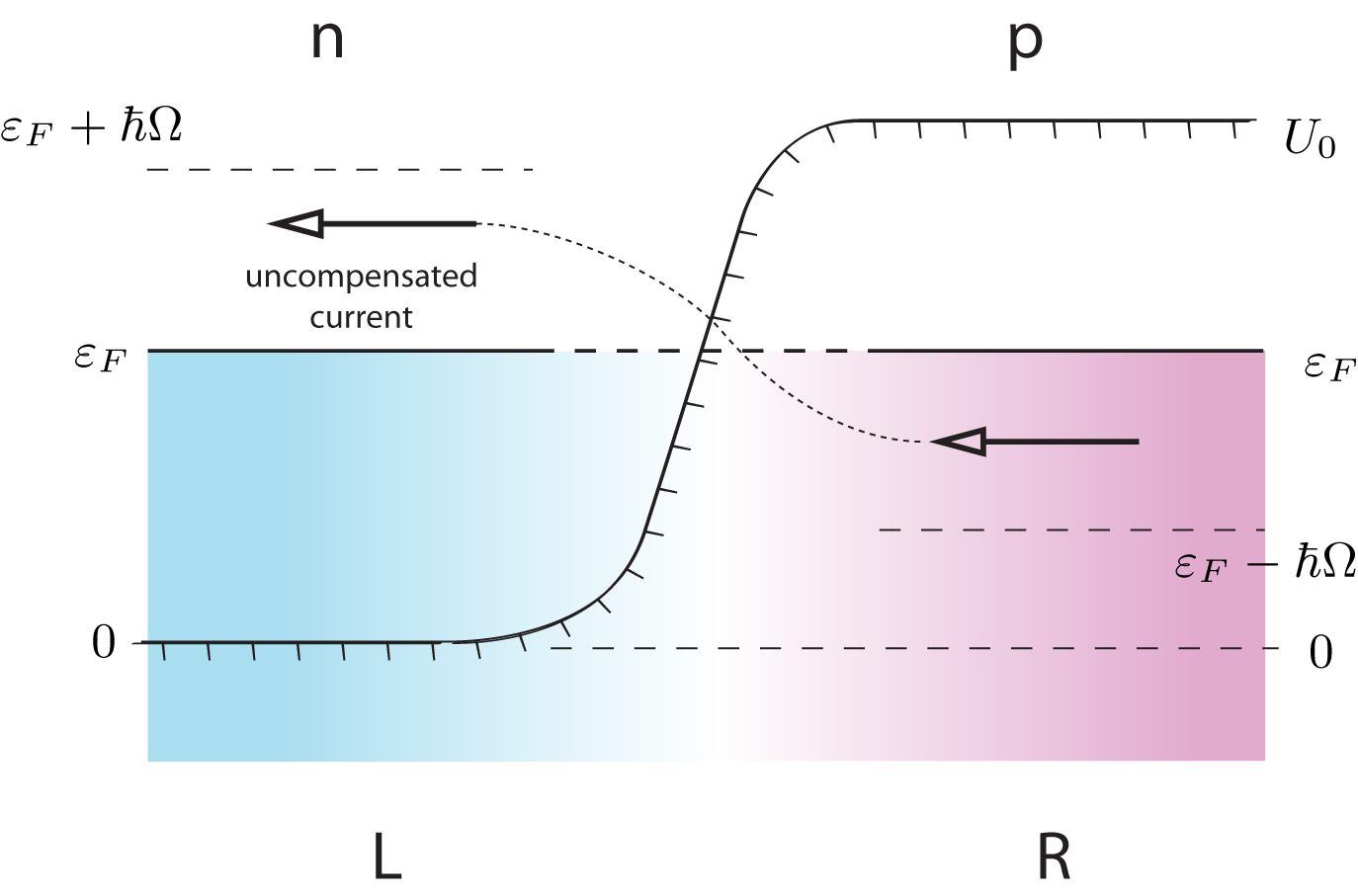}
 \caption{\label{fig:pntwo} (Color online) Contribution of electrons with different energies
 into the photocurrent through graphene p-n junction.}
\end{figure}

This can be understood as follows. Each electron with energy
$\varepsilon_F-\hbar\Omega<\varepsilon^\prime<\varepsilon_F$ in the
right lead penetrates into the left lead increasing its energy by
$\hbar\Omega$ (Fig.~\ref{fig:pntwo}). The current, carried by these
electrons, is not compensated by the corresponding time-reversed
processes, since the states of electrons, incoming from the left
lead on the energies
$\varepsilon_F<\varepsilon<\varepsilon_F+\hbar\Omega$ above the
Fermi level, are unoccupied. So, the electrons coming from the leads
on the correspondingly lower energies either compensate each others'
contributions to the currents or return back to their initial leads.

Thus, in order to find the photocurrent in the junction, one has to
apply Eq.~(\ref{moskalets}), performing integration over the states
of the left lead with energies in the interval $(\varepsilon_F;
\varepsilon_F+\hbar\Omega)$.

Let us assume that at some place around the resonant points the
potential has a constant slope $F$. Then, according to the results
of Section~\ref{sec:tunneling}, the probability for an electron at
some energy $\varepsilon$ and angle of incidence $\theta$ to go
along the trajectory under consideration, is
\begin{eqnarray}
    P_{LR}(\varepsilon,\varepsilon-\hbar\Omega,\theta)\nonumber\\
    =T(1-T)=e^{-\cL\cos\gamma}\left(1-e^{-\cL\cos\gamma}\right),
\end{eqnarray}
where $\gamma$ is the angle between the electron momentum at the
resonant point and the normal to the junction [cf.
Figs.~\ref{fig:proc1}(b) and \ref{fig:proc2}(b)], $\cL$ is the
parameter introduced by Eq.~(\ref{L}).

According to Eq. (\ref{moskalets}) the photocurrent is
\begin{eqnarray}
    I_{2D}=4W\int_{\varepsilon_F/v}^{(\varepsilon_F+{\hbar\Omega})/{v}}
    dp\int_{-\arcsin\frac{\hbar\Omega}{2pv}}^{\arcsin\frac{\hbar\Omega}{2pv}}
    \frac{p\:d\theta}{(2\pi\hbar)^2}ev\cos\theta
    \nonumber\\
    \times
2e^{-\cL\cos\gamma(p,\theta)}\left(1-e^{-\cL\cos\gamma(p,\theta)}\right).
\end{eqnarray}
The angle $\gamma$ can be determined from the transverse momentum
conservation law
\begin{eqnarray}
    p\sin\theta=\frac{\hbar\Omega}{2v}\sin\gamma.
\end{eqnarray}

Then
\begin{eqnarray}
    I_{2D}=
    4eW\int_{\varepsilon_F/v}^{(\varepsilon_F+{\hbar\Omega})/{v}}dp
    \int_{-\pi/2}^{\pi/2}d\gamma\frac{\hbar\Omega}{(2\pi\hbar)^2}\cos\gamma
    \nonumber\\
    \times
    e^{-\cL\cos\gamma}\left(1-e^{-\cL\cos\gamma}\right)\\
    =\frac{eW\Omega^2}{\pi
    v}\left(L_1(\cL)-I_1(\cL)-L_1(2\cL)+I_1(2\cL)\right),
    \label{photocurrent2D}
\end{eqnarray}
where $L_1(z)$ and $I_1(z)$ are correspondingly the modified Struve
and the modified Bessel functions of the first order
\cite{AbramowitzStegun}.

\begin{figure}[ht]
 \includegraphics[width=0.6\columnwidth]{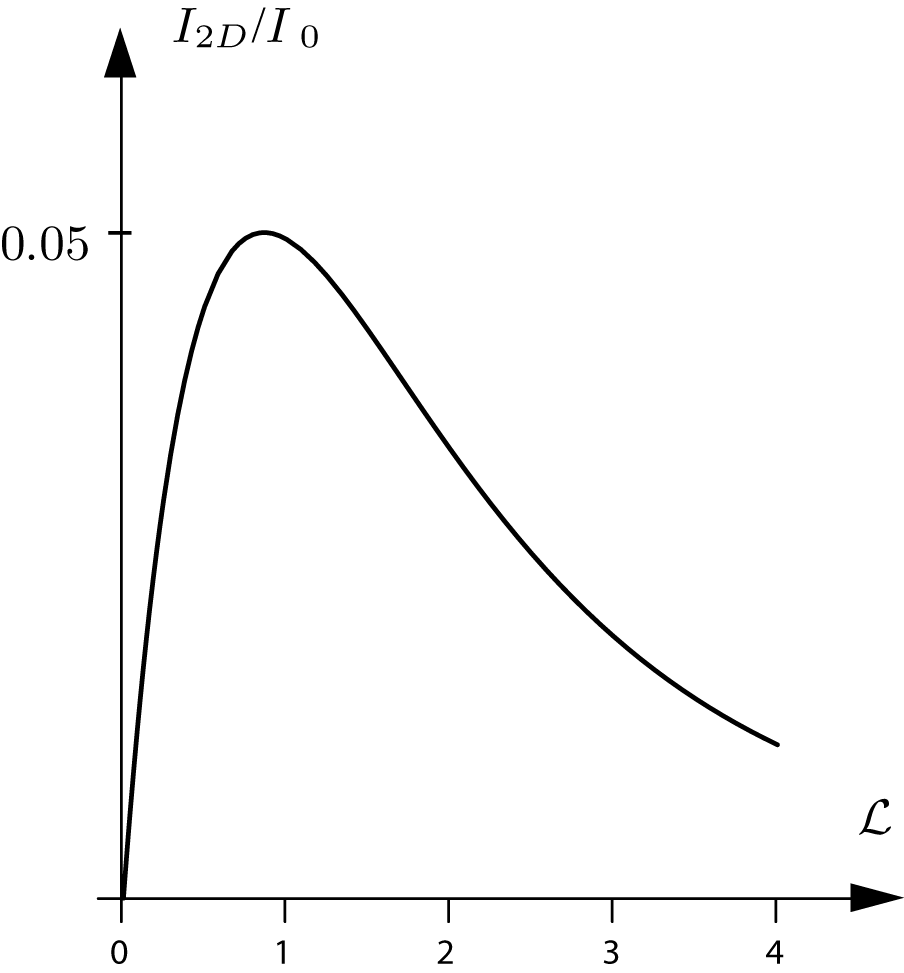}
 \caption{\label{fig:Photocurrent} Photocurrent in a graphene p-n junction as a function
 of the radiation intensity.}
\end{figure}

According to the Eqs.~(\ref{L}) and (\ref{Delta}), the parameter
$\cL$ is proportional to the intensity $S$ of the EF:
\begin{eqnarray}
    \cL=\frac{\pi^2e^2v}{\hbar cF}\frac{S}{\Omega^2}.
\end{eqnarray}
Expression (\ref{photocurrent2D}) is plotted in
Fig.~\ref{fig:Photocurrent},
\begin{eqnarray}
    I_0=\frac{eW\Omega^2}{v}.
\end{eqnarray}

{\it Small-intensity regime.}

In the experimentally relevant regime of small radiation
intensities ($\cL\ll1$) the photocurrent is
\begin{eqnarray}\label{photocurrentsmall}
    I_{2D}^{small}=\frac{eW\Omega^2}{2\pi v}\cL.
\end{eqnarray}
The last formula can be understood as follows. The effective
number of conducting ``2D-modes'', i.e. of transverse channels in
the energy interval $\sim\hbar\Omega$, is
\begin{eqnarray}\label{N2D}
    N_{2D}=W\frac{\Omega}{v}.
\end{eqnarray}
The probability to tunnel in each channel, i.e., along each
trajectory in the Figs.~\ref{fig:proc1}(b) and \ref{fig:proc2}(b),
is $\cL$, in the limit of small $\cL$. Then the Landauer-type
conductance of each channel is $G=\frac{e^2}{h}\cL$. Since each
electron acquires the energy $\hbar\Omega$ tunneling from right to
left, the effect of radiation on the conducting channels is
equivalent to the effective voltage $V_{eff}=\hbar\Omega/e$ applied
to the junction. Then Eq.~(\ref{photocurrentsmall}) transforms into
\begin{eqnarray}
    I_{2D}^{small}=N_{2D}GV_{eff}.
\end{eqnarray}

{\it Large-intensity regime.}

In the limit of large radiation powers ($\cL\gg1$)
Eq.~(\ref{photocurrent2D}) reduces to
\begin{eqnarray}\label{photolarge}
    I_{2D}^{large}=\frac{3eW\Omega^2}{4\pi^2v}\frac{1}{\cL^2}.
\end{eqnarray}
The photocurrent due to the modes under consideration is being
suppressed by strong radiation, since the electrons tend to be
reflected at each resonant point when $\cL$ is large and the
tunneling along the paths in the Figs.~\ref{fig:proc1}(b) and
\ref{fig:proc2}(b) becomes unlikely.

Since the photocurrent carried by electrons with sufficiently large
transverse momenta vanishes at $\cL\rightarrow\infty$, one needs to
consider the contribution of the other electrons incident almost
normally at the p-n interface.

\subsection{Photocurrent due to normal modes}

In the normal modes electrons have very small transverse momenta
$p_\bot$, so the problem can be viewed as one-dimensional. Introduce
the effective number of normal modes, i.e. of the channels where
electrons propagate without reflection at the p-n interface (taking
into account spin and valley degeneracies):
\begin{eqnarray}\label{N1D}
    N_{1D}=4\frac{W}{2\pi\hbar}\int_{-\infty}^{\infty}dp_\bot\:
    e^{-\frac{\pi p_\bot^2v}{\hbar F_0}}=
    \frac{2W}{\pi\hbar}\sqrt{\frac{\hbar F_0}{v}}.
\end{eqnarray}
As discussed before, when the condition (\ref{largepf}) is
satisfied, one has the inequality $N_{2D}\gg N_{1D}$.

The corresponding effective one-dimensional density of states (per a
unit of longitudinal length) is
\begin{eqnarray}
    \nu_{1D}=\frac{N_{1D}}{2\pi\hbar v}.
\end{eqnarray}

The current, flowing through the junction within the one-dimensional
picture, is \cite{Moskalets:cool} [cf. also Eq.~(\ref{moskalets})]
\begin{eqnarray}
    I=\int d\varepsilon\:\nu_{1D}\sum_n P_{LR}(\varepsilon,\varepsilon+n\hbar\Omega)
    \nonumber\\
    \times
    \left(f_L^{in}(\varepsilon)-f_{R}^{in}(\varepsilon+n\hbar\Omega)\right),
\end{eqnarray}
where $P_{LR}(\varepsilon,\varepsilon+n\hbar\Omega)$ is the
probability to tunnel from left to right from the state with energy
$\varepsilon$ to the one with $\varepsilon+n\hbar\Omega$.

The only tunneling process contributing to the photocurrent [Fig.
\ref{fig:resnormal}(b)] is the one when the particle is being
reflected at both resonant points, emitting two photons. The
probability of this double reflection is $(1-e^{-\cL})^2$.

The photocurrent due to the normal modes is possible provided
$U_0>\hbar\Omega$. Again, we assume that the height of the potential
barrier and the Fermi level are large enough, so that the
photocurrent due to the processes under consideration not simply
exists, but also reaches maximum as a function of these parameters,
\begin{eqnarray}
    \varepsilon_F>\frac{3\hbar\Omega}{2},\\
    U_0-\varepsilon_F>\frac{3\hbar\Omega}{2},
\end{eqnarray}
so that both resonant points are present on the potential for each
electron in the energy interval of the width $2\hbar\Omega$.

Then the photocurrent due to the normal modes is
\begin{eqnarray}\label{photocurrent1D}
    I_{1D}=\frac{e^2}{h}\left(1-e^{-\cL}\right)^2\left(\frac{2\hbar\Omega}{e}\right)N_{1D}.
\end{eqnarray}
We see thus, that at very large intensities ($\cL\gg1$) the
photocurrent is saturated. This happens because in this limit any
electron, reflected twice by the radiation with the probability
close to $1$, recovers its initial direction of velocity and
penetrates into the opposite lead.

\subsection{Full current vs. radiation intensity}

To sum up, the full photocurrent $I$ flowing through the irradiated
p-n junction is given by the sum
\begin{eqnarray}
    I=I_{2D}+I_{1D}
\end{eqnarray}
with $I_{2D}$ and $I_{1D}$ given by Eqs.~(\ref{photocurrent2D}) and
(\ref{photocurrent1D}) respectively.

 If the radiation power is
not too large,
\begin{eqnarray}
    \cL^2\lesssim\frac{3}{4\pi}\frac{N_{2D}}{N_{1D}},
\end{eqnarray}
the p-n junction is effectively two-dimensional, and the
photocurrent in it is described by Eq.~(\ref{photocurrent2D}). For
larger intensities, when the last inequality reverts, the
photocurrent strongly decreases (by the factor of $\sim
N_{2D}/N_{1D}$) and saturates at the value
$I_{1D}(\cL\rightarrow\infty)=e\hbar\Omega N_{1D}/\pi$.

\subsection{Suppression of the tunneling}

{\it Current-voltage characteristic of a strongly irradiated
junction.} Assume now that the voltage $V$ is applied to the
junction. Here we make the convention that the voltage is positive
in the case of the forward bias, when the electric potential of the
p-type graphene is larger than that of the n-type. The corresponding
direction of the current in the absence of radiation is considered
positive.

Introducing the conductance $G_{ball}$ of the ballistic graphene p-n
junction
\begin{eqnarray}
    G_{ball}=\frac{e^2}{h}N_{1D},
\end{eqnarray}
 and taking
into account Eq.~(\ref{photocurrent1D}), the current-voltage
characteristic of the junction under a very strong irradiation
($\cL^2\gg N_{2D}/N_{1D}$) can be written as
\begin{eqnarray}
    I(V)=G_{ball}\left(V-\frac{2\hbar\Omega}{|e|}\right).
\end{eqnarray}
The second term in the parenthesis of the last equation is the
photocurrent, Eq.~(\ref{photocurrent1D}) written in the limit
$\cL\rightarrow\infty$. The equation holds until
$V<U_0-\varepsilon_F+\hbar\Omega/2$. For larger voltages the current
grows even more rapidly due to the appearance of normal modes
without resonant points.

Thus, the current through the p-n junction with high enough
potential barrier cannot be suppressed by the radiation, due to the
existence of the channels where electrons can tunnel with
probability $1$, gaining or losing two quanta $\hbar\Omega$. As a
result, the differential conductance of the junction is the same as
in the absence of radiation.

{\it Conditions for tunneling suppression.}

In order to suppress the tunneling in the junction one has to cancel
the photocurrent due to the ``1D modes'', i.e. to make a
sufficiently low potential barrier $U_0<\hbar\Omega$ (or increase
the frequency), such that the normal trajectories with two resonant
points no longer exist.

Making a p-n junction with $U_0>\hbar\Omega/2$ and with one resonant
point for electrons coming with the Fermi energy, and then applying
strong radiation one can exponentially suppress the conductance due
to the normal modes and suppress the photocurrent carried by the
``2D modes'', because it becomes proportional to $\propto\cL^{-2}$,
Eq.~(\ref{photolarge}). One may also think of the situation for
$U_0<\hbar\Omega/2$ when one resonant point still exists for
electrons coming at the Fermi energy. In that case, applying a small
voltage $V$ to the junction, one obtains the normal current $I\sim
I_{2D}^{large}(eV/\hbar\Omega)\sim(e^2W\Omega V/v\hbar)\cL^{-2}$,
which is also suppressed at large intensities as $\propto\cL^{-2}$.

\section{Photocurrent in disordered samples}

\label{sec:disorder}

In the previous sections we considered purely ballistic
quasiparticle transport in irradiated graphene samples. In
principle, the transport can be strongly affected by disorder,
electron-phonon and electron-electron interactions. According to the
Ref.~\onlinecite{Fogler:nonball}, as concerns their conductance,
junctions fabricated in recent experiments are at the best in the
crossover between the ballistic and diffusive regimes.

In this section we calculate the photocurrent in a p-n junction
under a weak irradiation, $\cL\ll1$, in a disordered sample in
presence of electron-electron interaction. We show that the
photocurrent is the same as previously calculated in a ballistic
sample, Eq.~(\ref{photocurrentsmall}), provided that the
impurities-induced resistance of some resonant region in the
junction, where the carriers are effectively excited by the
radiation, is not large compared to the ballistic resistance of the
junction.

In this section we use the following assumptions. We neglect the
relaxation due to electron-phonon interaction on the length of the
junction and assume that the condition (\ref{indcondition}) is
fulfilled. The validity of such approximations will be confirmed in
Sec.~\ref{sec:experiment} by explicit estimates. For simplicity, we
suppose that the intervalley scattering is weak and neglect it.
Taking this scattering into account is possible in a similar way,
which leads to the similar formulas.
When writing the kinetic equations we also neglect the possibility
of disorder-assisted photon emission or absorption. In fact,
electrons can rebound from the impurities, gaining or losing energy
quanta $\hbar\Omega$ in the same way as from the smooth potential
$U(\br)$, considered in Sec.~\ref{sec:tunneling}. Such a process
would open additional channels for the tunneling of electrons from
one lead into the another and thus would increase the photocurrent.

Introduce the distribution function of electrons in the conduction
band, $f_\uparrow(\bp,\br,t)$, and in the valence band,
$f_\downarrow(\bp,\br,t)$. The first ones have their momenta
parallel to their pseudospins, while the second- antiparallel. The
stationary distributions are described by the kinetic equations (for
the derivation see the Appendix)
\begin{eqnarray}
    \left(-\partial_\br                         \label{kinmain1}
    U\partial_\bp+v\frac{\bp}{p}\partial_\br\right)f_\uparrow
    =\Gamma(\bp)(f_\downarrow-f_\uparrow)+(\St f)_\uparrow,\\
    \left(-\partial_\br                         \label{kinmain2}
    U\partial_\bp-v\frac{\bp}{p}\partial_\br\right)f_\downarrow
    =\Gamma(\bp)(f_\uparrow-f_\downarrow)+(\St f)_\downarrow,
\end{eqnarray}
where  $\Gamma(\bp)$ is the rate of the radiation-induced
pseudospin-flips, and
\begin{eqnarray}
    \St f=\left(\St f\right)^{imp}+\left(\St f\right)^{ee}
\end{eqnarray}
is the collision integral for impurity scattering and
electron-electron interaction.

Kinetic equations (\ref{kinmain1}), (\ref{kinmain2}) are valid in
the case of weak enough radiation, $\cL\ll1$. The effect of the
radiation and the effect of other processes on the electron dynamics
are described by two independent terms in the right-hand sides
(rhhs) of the equations. The mutual influence of the two
corresponding transition rates on each other can be neglected due to
the fulfillment of the condition (\ref{indcondition}), derived in
Sec.~\ref{sec:tunneling} from the considerations of Landau-Zener
tunneling in the momentum space. The same condition,
[\ref{indcondition2}], follows from the explicit derivation of the
kinetic equations (cf. Appendix). Let us emphasize, that treating
perturbatively the effect of disorder and radiation-induced
transitions in Eqs.~(\ref{kinmain1}) and (\ref{kinmain2}) one should
use the values of velocity $v$ and collisional terms, renormalized
by disorder \cite{Aleiner:renorm} and electron-electron interactions
\cite{Mishchenko:ee}.

The contribution of elastic impurities into the collision integral
has the form
\begin{eqnarray}\label{collintegral}
    \left(\St f_{\uparrow,\downarrow}(\bp,\br)\right)^{imp}\nonumber\\
    =\int d\bp^\prime
    w_{\bp\bp^\prime}(f_{\uparrow,\downarrow}(\bp^\prime,\br)-f_{\uparrow,\downarrow}(\bp,\br))
    \delta(p-p^\prime),
\end{eqnarray}
where the quantity $w_{\bp\bp^\prime}$ satisfies the condition
following from the time-reversal symmetry
\begin{eqnarray}\label{wtr}
    w_{\bp\bp^\prime}=w_{\bp^\prime\bp},
\end{eqnarray}
similarly to the analogous Eq.~(\ref{ttr}) in
Sec.~\ref{sec:genform}.

Analogously one can write the collision integral $\left(\St
f\right)^{ee}$ for the electron-electron interaction (see, for
instance, Ref.~\onlinecite{Mishchenko:ee}). However, further we will
not need an explicit form of this integral using instead only the
fact that the electron-electron interaction does not considerably
modify the resistance of the junction, i.e. that such a modification
is much smaller than the ballistic resistance of the junction.

We derive the photocurrent from the kinetic equations
(\ref{kinmain1}), (\ref{kinmain2}), making perturbation theory in
$\Gamma(\bp)$. In the zeroth order, when $\Gamma(\bp)=0$, neglecting
the change of the distribution functions due to the collision
integral on the energy scales of order of $\varepsilon_F$ and
$\hbar\Omega$, we can write the solution of Eq.~(\ref{kinmain1}) as
an equilibrium Fermi distribution
$f^0_\uparrow(\bp,\br,t)=\theta(\varepsilon_F-vp-U(\br))$, where
$\theta$ is the theta-function. Similarly, the solution of
Eq.~(\ref{kinmain2}) in absence of radiation is
$f^0_\downarrow(\bp,\br,t)=\theta(\varepsilon_F+vp-U(\br))$.

\begin{figure}[ht]
 \includegraphics[width=2in]{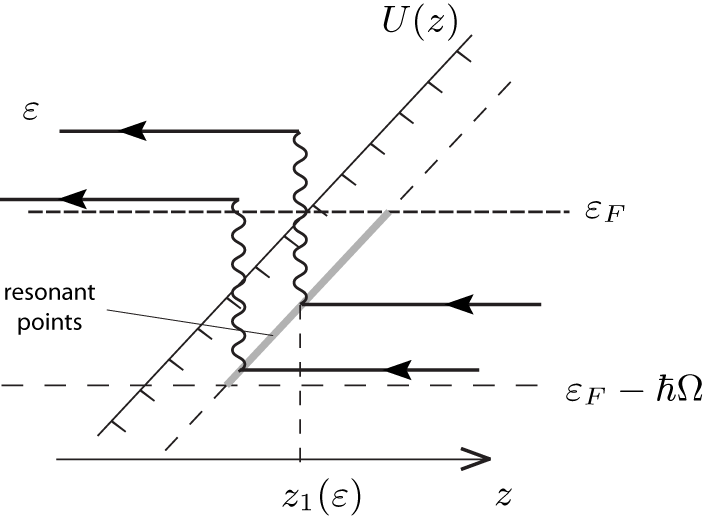}
 \caption{\label{fig:depletion} Radiation-induced excitation processes, which contribute
 to the photocurrent in the graphene p-n junction.
 }
\end{figure}

Taking into account the processes of the first order in
$\Gamma(\bp)$, one arrives at a slightly modified version of these
distributions; the radiation excites some small number of electrons
with energies
$\varepsilon_F-\hbar\Omega<\varepsilon^\prime<\varepsilon_F$ into
the energy interval
$\varepsilon_F<\varepsilon<\varepsilon_F+\hbar\Omega$. This happens
sufficiently close to the p-n interface, at
$|U(z)-\varepsilon_F|<\hbar\Omega/2$ (see Fig.~\ref{fig:depletion}).

Since above the Fermi energy the distribution function $f$ is small,
$f\ll1$, and below- $f\approx1$, we can rewrite Eq.~(\ref{kinmain1})
for the electrons in the valence band in the aforementioned region
as
\begin{eqnarray}\label{kinsimple}
    \left(-\partial_\br
    U\partial_\bp+v\frac{\bp}{p}\partial_\br\right)f_\uparrow
    =\Gamma(\bp)+(\St f)_\uparrow.
\end{eqnarray}

Let us multiply this equation by the factor $4pe\:d\varepsilon
d\theta/(h^2v)$ and integrate over the angle $\theta$ or, in other
words, over the direction of momentum $\bp$. Introduce
\begin{eqnarray}
    \overline\Gamma=\frac{\pi\Delta^2}{\hbar}\delta(\hbar\Omega-2pv),
\end{eqnarray}
which is the rate $\Gamma(\bp)$, averaged over the direction of
$\bp$, and $\nu_0=\Omega/(\pi\hbar v^2)$, the density of states per
unit square at the resonant point (the latter takes into account
spin and valley degeneracies). From Eq.~(\ref{kinsimple}) we get
after the integration
\begin{eqnarray}\label{chargecont}
    \diverg\left(\frac{\partial\bj_\uparrow(\varepsilon,z)}{\partial\varepsilon}d\varepsilon\right)=
    {e\overline{\Gamma}}{\nu_0}\:d\varepsilon
    +e\left(\frac{\partial}{\partial t}\frac{\partial n_\uparrow}{\partial \varepsilon}\right)^{ee}d\varepsilon,
\end{eqnarray}
where
\begin{eqnarray}
    \frac{\partial n_\uparrow(\varepsilon,z)}{\partial\varepsilon}d\varepsilon=
    d\varepsilon\cdot4\int\frac{p\:d\theta}{(2\pi\hbar)^2v}f_\uparrow(t,\bp,z)
\end{eqnarray}
and
\begin{eqnarray}
    \frac{\partial j_\uparrow(\varepsilon,z)}{\partial\varepsilon}d\varepsilon=d\varepsilon\cdot4e\int
    \frac{\bp\:d\theta}{(2\pi\hbar)^2} f_\uparrow(t,\bp,z)
\end{eqnarray}
are correspondingly the density of electrons and the density of the
current, carried by electrons with energies in the interval
$(\varepsilon,\varepsilon+d\varepsilon)$ in the conduction band at
the coordinate point $z$. The last term in Eq.~(\ref{chargecont})
describes the change of the density of electrons due to
electron-electron interactions.

Note, that due to the condition (\ref{wtr}), the contribution
(\ref{collintegral}) of elastic impurities into the collision
integral disappears in Eq.~(\ref{chargecont}) after the integration
over the direction of $\bp$. Indeed, Eq.~(\ref{chargecont}) is the
charge continuity equation for the carriers with energy in the
interval $d\varepsilon$, and the elastic scatterers cannot affect
the corresponding charge density. The two terms in the right-hand
side of Eq.~(\ref{chargecont}) represent the two inelastic processes
changing the density $(\partial n_\uparrow/\partial
\varepsilon)d\varepsilon$: external radiation and electron-electron
interaction.

The total current through the junction is given by the integral of
the current density over the energy:
\begin{eqnarray}\label{photocurrentpre}
    I_{ph}=W\int_{\varepsilon_F-\hbar\Omega}^{\varepsilon_F+\hbar\Omega}
    \left(\frac{\partial j_\uparrow(\varepsilon,z)}{\partial\varepsilon}+
    \frac{\partial j_\downarrow(\varepsilon,z)}{\partial\varepsilon}\right)d\varepsilon.
\end{eqnarray}
Depending on the coordinate point $z$, it can be carried either by
the particles in the conduction band or in the valence band or both.

When the junction is non-ballistic, the photocurrent depends on the
resistances of different parts of the junction and of the leads. Let
us calculate the photocurrent, assuming that the impurities are
present only in the ``resonant region'',
$|U(z)-\varepsilon_F|<\hbar\Omega/2$, while outside this region they
are absent, and the transport is purely ballistic. The parts of the
junction outside this region in the leads can be considered as some
external circuit, the resistance of which can be easily taken into
account after we obtain the final result for the photocurrent.

Integrating Eq. (\ref{chargecont}) over the longitudinal
coordinate $z$, we obtain
\begin{eqnarray}\label{jcurrent}
    \frac{\partial j_\uparrow(\varepsilon,z_1+0)}{\partial\varepsilon}-\frac{\partial j_\uparrow(\varepsilon,z_1-0)}{\partial\varepsilon}
    =\frac{1}{2}\cL ev\nu_0,\\
    \frac{\partial j_\uparrow(\varepsilon,z)}{\partial\varepsilon}=
    e\int_{z_1(\varepsilon)}^{z}\left(\frac{\partial}{\partial t}\frac{\partial n_\uparrow}{\partial
    \varepsilon}\right)^{ee} dz+const_1,\nonumber\\ \quad z>z_1,\\
    \frac{\partial j_\uparrow(\varepsilon,z)}{\partial\varepsilon}=
    e\int_{z_1(\varepsilon)}^{z}\left(\frac{\partial}{\partial t}\frac{\partial n_\uparrow}{\partial
    \varepsilon}\right)^{ee}  dz+const_2,\nonumber\\ \quad z<z_1,
\end{eqnarray}
where $z_1(\varepsilon)$ is the resonant point, corresponding to
the energy $\varepsilon$.

Electron-electron collisions conserve the total density of particles
$n(z)=\int(\partial n_\uparrow/\partial\varepsilon+\partial
n_\downarrow/\partial\varepsilon)d\varepsilon$ at a given point $z$,
which leads to the relation
\begin{eqnarray}\label{eepartconserv}
    \int d\varepsilon \left[\left(\frac{\partial}{\partial t}\frac{\partial n_\uparrow}{\partial
    \varepsilon}\right)^{ee}+\left(\frac{\partial}{\partial t}\frac{\partial n_\downarrow}{\partial
    \varepsilon}\right)^{ee}\right]=0.
\end{eqnarray}

If the ballistic resistance of the junction is much larger than the
characteristic diffusive resistance induced by impurities and
electron-electron interaction in the resonant region,
\begin{eqnarray}\label{diffllball}
    R_{ball}\gg R_{diff},
\end{eqnarray}
one can show, that the transmission of particles with energy
$\varepsilon$ through the point $z_0(\varepsilon)$, such that
$U(z_0)=\varepsilon$, is still strongly impeded, and the
photocurrent (\ref{photocurrentpre}) in the conduction and the
valence band is given mainly by the currents
$\frac{j_\uparrow(\varepsilon,z_1-0)}{\partial\varepsilon}d\varepsilon$
and $\frac{\partial
j_\downarrow(z_1+0)}{\partial\varepsilon}d\varepsilon$, as shown in
Fig.~\ref{fig:depletion}.

Indeed, if $dn_\uparrow$ is the density of excited electrons in the
energy interval $d\varepsilon$ in the resonant region, then, for
instance, the current density
\begin{eqnarray}
   \frac{\partial
   j_\uparrow(\varepsilon,z_1-0)}{\partial\varepsilon}\sim
   \frac{\partial n_\uparrow/\partial\varepsilon}{eWR_{diff}\nu_0}
\end{eqnarray}
 is much larger than
\begin{eqnarray}
    \frac{\partial
    j_\uparrow(\varepsilon,z_1+0)}{\partial\varepsilon}\sim
    \frac{\partial
n_\uparrow/\partial\varepsilon}{eW(R_{diff}+R_{ball})\nu_0},
\end{eqnarray}
 since
the condition (\ref{diffllball}) is fulfilled.

Taking into account the condition
\begin{eqnarray}
    \frac{\partial
    j_\uparrow(\varepsilon,z_1-0)}{\partial\varepsilon}\gg
    \frac{\partial
    j_\uparrow(\varepsilon,z_1+0)}{\partial\varepsilon}
\end{eqnarray}
and Eq.~(\ref{eepartconserv}), one can find the photocurrent by
integrating the current density at some point $z$ on the left from
the resonant point:
\begin{eqnarray}
     I_{ph}\approx W\int_{\varepsilon_F}^{\varepsilon_F+\hbar\Omega}
     \frac{\partial
     j_\uparrow(\varepsilon,z)}{\partial\varepsilon}d\varepsilon\\=
     \frac{1}{2}\cL ev\nu_0\hbar\Omega=\frac{eW\Omega^2}{2\pi
     v}\cL,\label{photocurrdis}
\end{eqnarray}
the same result as given by the Eq.~(\ref{photocurrentsmall}).

Thus, we have shown that in a disordered sample, where the ballistic
considerations of the previous sections cannot be immediately
applied due to the presence of elastic impurities and
electron-electron interaction in the resonant region, the
photocurrent does not change until the disorder or the interactions
become too strong, so that the impurities-induced or the
interactions-induced resistance becomes larger than the ballistic
one.

In the opposite limit, $R_{diff}\gg R_{ball}$, the excited electrons
and holes, created in the resonant region by the radiation, diffuse
almost independently of the external potential at the p-n interface.
The effect of large diffusive resistance of the resonant region on
the photocurrent is analogous to the effect of a large resistance of
the external circuit. The photocurrent in this regime is reduced as
\begin{eqnarray}
    I_{ph}^{diff}\sim\frac{R_{ball}}{R_{diff}}I_{ph},
\end{eqnarray}
with $I_{ph}$ given by Eq.~(\ref{photocurrdis}).

According to the Ref.~\onlinecite{Fogler:nonball}, in the recent
experiments \cite{Huard:stanfordexp, Williams:harvardpn} the
ballistic resistance is of the same order of magnitude as the
diffusive one, defined as a difference between the resistances of
n-n and p-n junctions. If $R_{ball}\sim R_{diff}$, Eq.
(\ref{photocurrdis}) is only an order-of-magnitude estimate of the
photocurrent. As follows from the derivation of the photocurrent,
the increase of the resistance $R_{diff}$ leads to the decrease of
the photocurrent. One can make the resistance $R_{diff}$ smaller
than the ballistic resistance, decreasing the frequency of the EF
and thus reducing the size of the resonant region and $R_{diff}$.

\section{Possibility of experimental observation of photocurrent}

\label{sec:experiment}

In this section we address the question of experimental
observability of the photocurrent in a graphene p-n junction. We
analyze the necessary conditions and estimate the value of the
photocurrent for achievable radiation intensities and the junction
parameters.

{\it Geometrical parameters and gate voltages.} As discussed in the
previous section, for the largest photocurrent one needs the
diffusive resistance of the junction to be smaller than the
ballistic one. This can be achieved by using sufficiently short
junctions \cite{Fogler:nonball}. Let us take the length of the
junction $L=100nm$, close to that in the experiment in
Ref.~\onlinecite{Huard:stanfordexp}, where the resistance of the
junction is described by the ballistic model rather than by the
diffusive one. The typical width $W$ of a p-n junction
\cite{Lemme:fet, Huard:stanfordexp, Williams:harvardpn,
Ozyilmaz:columbiapn} is a few micrometers; for our estimates we take
$W=1\mu m$.

Let $U_0=0.4eV$ and $\varepsilon_F=U_0/2$ be, respectively, the
height of the potential barrier and the Fermi energy, close to the
typical experimental parameters. For the slope of the potential we
use the naive estimate $F=U_0/L$, regarding it as constant along the
junction. To be more precise, considering the effective potential,
one should take into account the non-uniform charge density
distribution in the junction \cite{ZhangFogler}. However, the
corrected in such a way potential profile would have the slope of
the same order of magnitude as the naive estimate.

{\it Characteristic relaxation lengths.}

For the chosen parameters of a junction the characteristic length of
Landau-Zener tunneling, given by Eq.~(\ref{rLZ}), is
\begin{eqnarray}
    \vartriangle r_{LZ}\sim\sqrt{\frac{\hbar v}{F}}\approx 13nm.
\end{eqnarray}
According to the experiment in Ref.~\onlinecite{Dawlaty:photoex},
where the relaxation time of carriers excited by the near-infrared
light has been measured, electron-electron interaction is
responsible for the fastest stage of relaxation, occurring on the
typical time $\sim0.1ps$, corresponding to the electron path
$l_{ee}\sim0.1\mu m$. Since $l_{ee}\gg\vartriangle r_{LZ}$, the
condition (\ref{indcondition}) is fulfilled, as we assumed in the
previous section. Another slower stage of relaxation due to the
electron-phonon interaction has a characteristic length $\sim1\mu
m$.

The mean free path of carriers in graphene is of order of $1\mu m$
at room temperature \cite{Novoselov:firstgraphene,
Morozov:twodeghall}. The characteristic length of relaxation due to
electron-phonon interaction should be of the same order or larger.
Then the neglect of such a relaxation in the previous section is
quite a reasonable approximation.

 {\it Desirable radiation frequency.}
Calculating the current due to ``2D-modes'' in the
Sec.~\ref{sec:photocurrent}, we dealt with momentum scales much
larger than $p_\bot^0=(\hbar F/\pi v)^\frac{1}{2}$. Accordingly, to
have resonant points on the electron trajectories in the
``2D-modes'', one should apply the radiation with angular frequency
$\Omega$ much larger than
\begin{eqnarray}
    \Omega_{ir}=\frac{p_\bot^0 v}{\hbar}=\left(\frac{vF}{\pi\hbar}\right)^\frac{1}{2}.
\end{eqnarray}
If $\Omega\lesssim\Omega_{ir}$, then the ``2D-modes'' and some
``1D-modes'' do not have resonant points, that is, there exist
electrons only weakly affected by the EF and freely penetrating
through the p-n interface without reflection. As a result, if the
frequency $\Omega$ is too low, the photocurrent is strongly reduced,
and the suppression of tunneling is impossible.

 Note, that according to the
Eqs.~(\ref{N1D}) and (\ref{N2D}) the condition
$\Omega\gg\Omega_{ir}$ is equivalent to $N_{2D}\gg N_{1D}$, ensuring
that the junction is effectively two-dimensional. Provided this
condition is fulfilled, the current is carried mainly by the
``2D-modes''. For our choice of junction parameters
$\hbar\Omega_{ir}\approx29meV$, which corresponds to the frequency
$f_{ir}=\Omega_{ir}/(2\pi)\approx7THz$.

As we noted in the Sec.~\ref{sec:photocurrent}, the photocurrent is
possible if $U_0>\hbar\Omega/2$, i.e. $\hbar\Omega<800meV$ (or
$\Omega<200THz$). However, to maximize the photocurrent one should
satisfy conditions (\ref{phmaxcond1}) and (\ref{phmaxcond2}), for
the case under consideration equivalent to
$\hbar\Omega<U_0/2=200meV$.

{\it Magnitude of the photocurrent.} The characteristic radiation
intensity used in the experiments with nanotube junctions
\cite{Freitag:photofet, Wang:plpc} is about a few $kW/cm^2$. Assume,
the same intensities can be applied to graphene junction, and set
$S=10kW/cm^2$, close to the maximal value reached in
Ref.~\onlinecite{Freitag:photofet}.

Then the photocurrent is
\begin{eqnarray}
    I=\frac{\pi e^3W}{2\hbar cF}S\approx0.3\mu A,
\end{eqnarray}
independently of the frequency in the desirable range
$\hbar\Omega_{ir}<\hbar\Omega<U_0/2$. Note, that the photocurrent is
a few orders of magnitude larger than those obtained in the
experiments with carbon nanotubes \cite{Freitag:photofet,
Ohno:photofet, Marcus:photofet}.

{\it Possibility of tunneling suppression.} To maximize the
dynamical gap, Eq.~(\ref{Delta}), one should lower the frequency.
For $\hbar\Omega=\hbar\Omega_{ir}=29meV$ we obtain the dynamical gap
$\Delta\approx6meV$ and the exponent of tunneling through it--
$\cL\approx10^{-3}$, which is insufficient to suppress the
tunneling. To confine electrons, i.e. to achieve $\cL\gg1$, one
should use proportionally larger radiation powers or longer
junctions.

\section{Conclusion}

To sum up, we studied electron transport in graphene junctions
irradiated by monochromatic electromagnetic field (EF).

 The radiation opens dynamical gaps in the quasiparticle spectra,
proportional to the amplitude of the EF and inversely proportional
to its frequency, Eq.~(\ref{Delta}). The appearance of the gaps
results in a strong modification of current-voltage characteristics
of a junction.

If the height of the potential barrier is large enough, the directed
current (\emph{photocurrent}), Eq.~(\ref{photocurrent2D}), flows
through the junction without any dc bias voltage applied. At small
radiation intensities, the photocurrent, proportional to the
radiation power, Eq.~(\ref{photocurrentsmall}), is a result of
inelastic quasiparticle tunneling assisted by one-photon absorption.
At large intensities, the photocurrent, Eq.~(\ref{photolarge}),
decreases with radiation power and finally saturates at some
constant value, Eq.~(\ref{photocurrent1D}).

When the potential barrier is smaller than the photon energy
$\hbar\Omega$ but larger than $\hbar\Omega/2$, the saturation does
not happen and the photocurrent decreases to zero at large radiation
intensities. When the potential barrier is smaller than
$\hbar\Omega/2$, any photocurrent is absent. In these regimes, one
can adjust the Fermi level in such a way, that the quasiparticle
transmission in the junction is determined by the tunneling through
the gap and can be fully suppressed, provided that the radiation
power is large enough.

In the present paper we also analyze the influence of elastic
impurities and electron-electron interaction on the magnitude of the
photocurrent, and show that they weakly affect the photocurrent, if
the diffusive resistance of the junction is not too large compared
to the ballistic one.

{\it Acknowledgements.} We thank L.I.~Glazman, M.Yu.~Kharitonov, and
A.F.~Volkov for useful discussions. This work has been financially
supported by SFB Transregio 12 and SFB 491.

\appendix

\section{Kinetic equation in irradiated graphene}

\label{sec:appendix}

Now we derive explicitly the kinetic equation, governing the
dynamics of the electron distribution functions in graphene exposed
to monochromatic electromagnetic wave, taking into account the
effect of disorder, electron-electron and electron-phonon
interactions on the transport.
 Since
graphene in the vicinity of some resonant point can be considered as
a semiconductor with the spectrum, linearized close to the resonant
momentum, the dynamics of carriers, for which the radiation matters,
should be the same as for conventional semiconductors
\cite{Henneberger:kinetics, Henneberger:kinetics2, Jahnke:kinetics}.

As the radiation can only flip the pseudospin and does not induce
intervalley scattering, deriving the kinetic equation in the lowest
non-vanishing order in the radiation power, we can limit ourselves
to the consideration of dynamics in a single valley and a single
spin direction, because the intervalley- and spin- scattering would
enter only the part of the collision integral, which is independent
of the radiation.

It is convenient to perform calculations in the basis of electron
states $|\uparrow_\bp\rangle$, pseudospin is directed along the
momentum $\bp$, and $|\downarrow_\bp\rangle$, pseudospin is
antiparallel to $\bp$. We choose correspondingly the coordinate
system in the momentum space such that $z$ axis is directed along
$\bp$ and the $x$ axis is perpendicular to $\bp$ and parallel to the
graphene plane (the plane, in which $\bp$ can vary), and the $y$
axis- normally to the plane. This frame fixes the pseudospin basis.

Since the basis depends explicitly on $\bp$, in the momentum
representation one should substitute in the Hamiltonian (\ref{Ham})
the operator of spatial coordinate $\hat\br$ by the covariant
momentum derivative:
\begin{eqnarray}
    \hat\br\rightarrow{\widetilde\br}=
    i\frac{\partial}{\partial\bp}+\frac{1}{2}\frac{\partial\alpha}{\partial\bp}\hsigma_y,
\end{eqnarray}
where $\alpha(\bp)$ is the angle of rotation of the frame about the
$y$ axis, normal to the graphene plane. In this section $\hbar=1$.
 The second term in the
last expression is the gauge potential due to the local frame
rotations in the momentum space.

The modulus of this term for a given momentum $\bp$ is of order of
the corresponding Fermi wavelength
$\lambda_F\sim(\partial\alpha/\partial p)$. Far from the Dirac point
it is much smaller than the characteristic scale on which the
potential $U(\br)$ changes. This allows us to expand the potential
up to the first order in the gauge field, and write the Hamiltonian
as
\begin{eqnarray}
    \cH\approx vp\hsigma_z+U(\br)+(2p)^{-1}{\partial_x U(\br)}\hsigma_y
\end{eqnarray}
in the basis chosen. Here $\partial_x U(\br,\bp)= (dU(\br)/d\br)
\be_x(\bp)$, $\be_x$ is the unit vector directed along the
perpendicular to momentum $\bp$ the $x$ axis in the graphene plane.

Now let us proceed to the derivation of the kinetic equation for the
distribution functions in the basis of states
$|\uparrow_\bp\rangle$, $|\downarrow_\bp\rangle$. Analogously to the
field operator $\hPsi=(\hPsi_e,\hPsi_h^\dagger)^T$
 in a conventional semiconductor
\cite{Henneberger:kinetics}, we introduce the operator
$\hPsi=(\hPsi_\uparrow,\hPsi_\downarrow)^T$ with two components
acting correspondingly in the conduction and the valence bands of
graphene. In the momentum representation the indices $\uparrow$ and
$\downarrow$ of the operator $\hPsi(\bp)$ refer to the particles
with pseudospins aligned along or opposite to the momentum $\bp$,
respectively. Then we introduce the non-equilibrium Green's
functions
\begin{eqnarray}
    G_{ab}^<(1,2)=i\langle \hPsi^\dagger_b(2),
    \hPsi_a(1)\rangle,\\
    G_{ab}^>(1,2)=-i\langle \hPsi_a(1),
    \hPsi_b^\dagger(2)\rangle,
\end{eqnarray}
where $a,b=\uparrow, \downarrow$; $1=\{t_1,\br_1\}$;
$2=\{t_2,\br_2\}$. 
 Accordingly, we define
\cite{RammerSmith} the $2\times 2$ matrix Green's functions $G^A$,
$G^R$, and $G^K$, and the matrix function
\begin{eqnarray}
    \underline G=\left(
    \begin{array}{cc}
    G^R & G^K \\
    0 & G^A
    \end{array}
    \right),
\end{eqnarray}
which satisfies the equation \cite{RammerSmith}
\begin{eqnarray}\label{Dyson}
   [(\underline G_0^{-1}-\underline\Sigma) \otimes \underline
   G]=0
\end{eqnarray}
(Dyson equation minus its conjugate), where
$G_0^{-1}(1,1^\prime)=(i\partial_{t_1}-\varepsilon(1))\delta(1-1^\prime)$,
$\varepsilon$ is the Hamiltonian of the particles unperturbed by the
radiation and impurities, and $\underline\Sigma$ is the self-energy.
The square brackets here stand for the commutator $[A\otimes
B]=A\otimes B-B\otimes A$.

Let us decompose $\underline\Sigma$ into two parts,
\begin{eqnarray}
    \underline\Sigma(1,1^\prime)=\hat
    V(t_1)\delta(1-1^\prime)+\underline\Sigma_i(1,1^\prime),
\end{eqnarray}
where $\hat V(t)$ is the EF-induced perturbation of the
single-particle Hamiltonian, $\underline\Sigma_i(1,1^\prime)$- the
rest of the self-energy part. Assuming that the EF is weak and
purposing to find the dynamics of the carries in the lowest order in
the radiation power, we will neglect the effect of external
radiation on $\underline\Sigma_i$.

It is convenient to solve the problem in the Wigner
representation, introducing the ``center of mass'' coordinates
$T=(t_1+t_2)/2$, $\bR=(\br_1+\br_2)/2$ and the relative ones,
$\tau=t_1-t_2$, $\br=\br_1-\br_2$, and making Fourier transform of
all the Green's functions with respect to $\tau$ and $\br$.

However, unlike the usual situation \cite{RammerSmith}, we expect
that in the Wigner representation only the quantities
\begin{eqnarray}
\underline
G_0^{-1}(T,\bR,\bp,\varepsilon)=\nonumber\\=\varepsilon-vp\hsigma_z-U(\bR)-(2p)^{-1}\partial_x
U(\bR,\bp)\hsigma_y,
\end{eqnarray}
$G_{aa}^K$, and $\underline\Sigma_i$ are the slow functions of time
$T$, i.e. vary on the time scales much larger than the inversed
relevant kinetic energies of electrons, while the other functions,
$G^{A/R}$ and $G_{ab}^K$, contain contributions proportional to the
fast in $T$ perturbation $\hat V(T)=-({ev}/{c})\bA(T)\bsigma$. The
diagonal elements $G_{aa}^K$ of the Keldysh Green's function should
be slow since they depend only on the electron distribution
function, which should vary slowly due to the weakness of
perturbation. Below we confirm this by the direct calculation.

In order to derive the kinetic equation, we consider the Keldysh
component of Eq.~(\ref{Dyson}):
\begin{eqnarray}
    [G_0^{-1}\otimes G^K]=[\hat V\otimes G^K]\nonumber\\
    +\Sigma_i^R\otimes G^K
    -G^K\otimes \Sigma_i^A
    +\Sigma_i^K\otimes G^A-G^R\otimes\Sigma_i^K.
\end{eqnarray}
Taking into account the slowness of $G^K$, one can rewrite the left
part of the last equation using the gradient approximation
\cite{RammerSmith}:
\begin{eqnarray}\label{semikinetic}
    i\bigg(\partial_T G^K-\partial_{\bR} U\partial_\bp G^K\nonumber
    \frac{1}{2}\left\{\frac{\bp}{p}v\hsigma_z,\partial_\bR G^K\right\}\\+i[vp\hsigma_z,G^K]
    +(2p^{-1})\partial_x U\left[\hsigma_y,G^K\right]\bigg)\nonumber\\
    =\left[\hat V\otimes G^K\right]
    +\Sigma_i^R\otimes G^K
    -G^K\otimes \Sigma_i^A\nonumber\\
    +\Sigma_i^K\otimes G^A-G^R\otimes\Sigma_i^K.
\end{eqnarray}
The brackets $[\:,]$ stand here for the commutators of matrices,
$\{\:,\}$- for the anticommutators. The left part of the last
equation describes the ballistic properties of electrons in
graphene. The last four terms describe the change of distribution
functions due to the electron scattering, independent of the
radiation. Further we do not consider these terms in detail and
focus on the radiation-induced transitions, i.e. on the first term
in the right-hand side of Eq.~(\ref{semikinetic}). We will only
assume, that the rate of the radiation-independent scattering
between the states $|\uparrow_\bp\rangle$ and
$|\downarrow_\bp\rangle$ is not too strong compared to the rate of
the radiation-induced transitions, so that the latter can be
calculated independently. The applicability of this assumption will
be discussed below.

The term $\left[({\partial_x U}/{2p})\hsigma_y,G^K\right]$ in the
left-hand side of Eq.~(\ref{semikinetic}) can be rewritten into the
right-hand side (rhs) as $-({\partial_x U}/{2p})[\hsigma_y\otimes
G^K]$ in the leading order in the gradient approximation. Then the
term $-({\partial_x U}/{2p})\hsigma_y$ can be considered in the
further calculations as an additional small perturbation, induced by
the electromagnetic field with frequency and amplitude going to zero
[cf. Eq.~(\ref{A})]. Such a perturbation could induce transitions
between electron states close to the Dirac point. However, here we
are interested in the radiation-induced transitions between the
states with large momenta close to the resonant ones. Then we will
disregard the last term in the left-hand side of
Eq.~(\ref{semikinetic}). Within such an approximation the kinetic
equation in the form (\ref{semikinetic}) in the ballistic graphene
coincides with that in a conventional semiconductor, as it should be
at large momenta.

Let us find the EF-induced modification of $G^K\equiv G^<+G^>$,
disregarding the other processes, the contribution of which into the
relaxation of the electron distribution, as we assumed, can be found
separately. Due to the radiation some non-zero off-diagonal terms
$G^K_{ab}$ appear. In the first order in perturbation in the
momentum-time representation we obtain

\begin{eqnarray}\label{Gless}
    G_{ab}^<(t_a,t_b)=i\langle\hPsi_b^\dagger(\bp,t_b)\hPsi_a(\bp,t_a)\rangle\nonumber
    \\ \approx
    -\int_0^{t_b}G_{aa}^<(t_a,\tau)V_{ab}(\tau)G_{bb}^>(\tau,t_b)d\tau
    \nonumber \\
    +\int_0^{t_a}G_{aa}^>(t_a,\tau)V_{ab}(\tau)G_{bb}^<(\tau,t_b)d\tau
    \nonumber \\
    +\int_{t_a}^{t_b}G_{aa}^<(t_a,\tau)V_{ab}(\tau)G_{bb}^<(\tau,t_b)d\tau.
\end{eqnarray}

From now on we assume that in the vicinity of the resonance under
consideration the RWA can be applied, and that the off-diagonal
elements of the perturbation are taken in the form
\begin{eqnarray}\label{perturbsf}
    V_{ab}(\tau)=W_{ab}e^{i\Omega_{ab}\tau}.
\end{eqnarray}

Then, from the Eq.~(\ref{Gless}) we find in the Wigner
representation
\begin{eqnarray}
    G_{ab}^<(T,\bR,\bp,\varepsilon)\approx\frac{2\pi
    W_{ab}e^{i\Omega_{ab}T}}{i(\varepsilon_a-\varepsilon_b+\Omega_{ab}-i0)}
    \nonumber
    \\ \times
    \left(f_b\delta\left(\varepsilon-\varepsilon_b+\frac{\Omega_{ab}}{2}\right)
    -f_a\delta\left(\varepsilon+\varepsilon_a-\frac{\Omega_{ab}}{2}\right)\right),
\end{eqnarray}
where we have introduced the total (kinetic+potential) energy
$\varepsilon_{a}$ of a particle with momentum $\bp$ and the
pseudospin $a$ and the distribution function
\begin{eqnarray}
    f_a(T,\bR,\bp)=\int \frac{d\varepsilon}{2\pi i}G^<_{aa}(T,\bR,\bp,\varepsilon).
\end{eqnarray}

Analogously,
\begin{eqnarray}
    G_{ab}^>(T,\bR,\bp,\varepsilon)\approx\frac{2\pi
    W_{ab}e^{i\Omega_{ab}T}}{i(\varepsilon_a-\varepsilon_b+\Omega_{ab}-i0)}
    \nonumber \\ \times
    \left((1-f_a)\delta\left(\varepsilon-\varepsilon_a-\frac{\Omega_{ab}}{2}\right)
    \right.\nonumber \\ \left.
    -(1-f_b)\delta\left(\varepsilon-\varepsilon_b+\frac{\Omega_{ab}}{2}\right)\right).
\end{eqnarray}

Now the obtained functions $G^>$ and $G^<$ can be used to calculate
the term $[\hat V\otimes G^K]$ in the rhs of
Eq.~(\ref{semikinetic}). The convolution of two functions in the
Wigner representation there can be obtained following the rules
formulated in the Ref.~\onlinecite{RammerSmith}. For instance,
\begin{eqnarray}
    V_{ba}\otimes
    G_{ab}^K=e^{-\frac{i}{2}\partial_T^V\partial_\varepsilon^G}V_{ba}G_{ab}
    \nonumber    \\
    =V_{ba}(T)G_{ab}^K\left(\varepsilon+\frac{\Omega_{ba}}{2},T\right).
\end{eqnarray}

Before substituting $G^K$ in Eq.~(\ref{semikinetic}) we integrate
over the energy $\varepsilon$ the diagonal element of this equation,
in order to arrive at the kinetic equation. In that connection we
also note, that the diagonal elements of the perturbation,
$V_{bb}(T)=\sum_\Omega W_{bb}(\Omega)e^{i\Omega T}$, do not change
the distribution functions $f_a$ and thus do not enter the kinetic
equation. This happens because the longitudinal perturbation cannot
change the energy level occupation. Indeed,
\begin{eqnarray}
    \int d\varepsilon\left[V\otimes G^K\right]_{bb}=\nonumber\\
    \sum_\Omega W_{bb}(\Omega)e^{i\Omega
    T}\int d\varepsilon\: G_{bb}^K\left(\varepsilon+\frac{\Omega}{2},T\right)-\nonumber\\
    -\sum_\Omega W_{bb}(\Omega)e^{i\Omega
    T}\int d\varepsilon\: G_{bb}^K\left(\varepsilon-\frac{\Omega}{2},T\right)=0.
\end{eqnarray}

Keeping the second-order terms in EF or the first-order ones in the
gradients of slow variables, we finally get the kinetic equation,
\begin{eqnarray}\label{kinetic0}
    \left(\partial_T-\partial_\bR
    U\partial_\bp+v(\hsigma_z)_{bb}\partial_\bR\right)f_b\nonumber\\
    =2\pi |W_{ba}|^2
    \delta(\varepsilon_a-\varepsilon_b+\Omega_{ab})(f_a-f_b)+
    (\St f)_b.
\end{eqnarray}
Here $(\St f)_b$ is the usual collision integral accounting for the
change of the distribution function $f_b$ irrespective of radiation
due to impurity scattering, electron-phonon and electron-electron
interactions. $(\sigma_z)_{bb}=\langle b\bp|\hsigma_z| b\bp\rangle$
is the direction of the particle velocity in the state with momentum
$\bp$ and pseudospin $b$: (along $\bp$ or opposite to $\bp$).

If we take the EF-induced perturbation in the form of
Eq.~(\ref{DeltaPert}) within the RWA in the vicinity of the
corresponding resonance the kinetics of, e.g., the function
$f_\uparrow(T,\bR,\bp)$ is described by the equation
\begin{eqnarray}\label{kinetic}
    \left(\partial_T-\partial_\bR
    U\partial_\bp+v\frac{\bp\partial_\bR}{p}\right)f_\uparrow\nonumber\\
    =\Gamma(\bp)(f_\downarrow-f_\uparrow)+
    (\St f)_b,
\end{eqnarray}
where
\begin{eqnarray}\label{Gamma}
    \Gamma(\bp)=\frac{2\pi}{\hbar} \Delta^2\sin^2(\widehat{\bp,\bE})\:
    \delta(2vp-\hbar\Omega),
\end{eqnarray}
where $\Delta$ is the dynamical gap, and the Planck constant is
recovered.

Equation (\ref{kinetic}) can be understood as follows. It differs
from the usual kinetic equation by the first term in the rhs
describing the rate of the EF-induced change of the distribution
function $f_\uparrow$. The change is due to the pseudospin flip
under irradiation: $|\uparrow\rangle\rightarrow|\downarrow\rangle$
and $|\downarrow\rangle\rightarrow|\uparrow\rangle$. The rate
$\Gamma(\bp)$ of both processes can be obtained from the Fermi's
golden rule. Note, that neither the distribution function
$f_\uparrow$ nor $f_\downarrow$ in Eq.~(\ref{kinetic}) is assumed to
be small; the Pauli exclusion principle is taken into account in the
adduced derivation or in the Fermi's golden rule, so the equation is
valid for arbitrary distribution functions.

Let us examine now the applicability of our kinetic equation.
Deriving the term accounting for the radiation-induced transitions,
the first term in the rhs of Eq.~(\ref{kinetic0}), we assumed that
the distribution functions $f_\uparrow$ and $f_\downarrow$ are
weakly perturbed by the radiation. As it follows from
Eqs.~(\ref{kinetic}) and (\ref{Gamma}), this condition is satisfied
when
\begin{eqnarray}
    \frac{\pi\Delta^2}{\hbar vF}=\cL\ll1.
\end{eqnarray}
Indeed, if we consider the normal incidence of particles on a smooth
potential barrier, using the obtained kinetic equation and setting
the collision integral equal to zero, we arrive at the tunneling
probability
\begin{eqnarray}
    T_{kinetic}=1-\cL,
\end{eqnarray}
which agrees with the result of Eq.~(\ref{T}) in the limit
$\cL\ll1$.

Another assumption we used when deriving the kinetic equation is
that the radiation-independent relaxation processes, such as
impurity scattering, electron-electron, and electron-phonon
interactions weakly influence the rate of the radiation-induced
transitions that can be calculated independently. If we took into
account this influence, the delta-function in Eq.~(\ref{kinetic0})
would have to be substituted by some function smooth on the scale
$\hbar/\tau_R$, where $\tau_R$ is some characteristic time of
relaxation due to the radiation-independent processes.

Then, according to Eq.~(\ref{Gamma}), weak radiation would affect
the distribution functions $f(T,\bR,\bp)$ in the momenta interval of
the characteristic width
\begin{eqnarray}
    \delta p\sim\frac{\hbar}{\tau_R v}
\end{eqnarray}
around the resonant $p_{res}=\hbar\Omega/(2v)$. For our
approximation to be valid, the distribution functions in this
momentum interval should be weakly changed by the
radiation-independent processes. As follows from
Eq.~(\ref{kinetic0}), the characteristic scale, on which momentum
relaxes due to these processes, is $\delta p_R\sim 1/(\tau_R F)$,
where $F$ is the characteristic slope of the potential $\partial_\bR
U(\bR)$ in the region under consideration. We need the fulfillment
of the condition $\delta p_R\gg \delta p$ and obtain thus
\begin{eqnarray}\label{indcondition2}
    \tau_R\gg\sqrt{\frac{\hbar}{vF}}.
\end{eqnarray}
Under this condition the radiation-induced transitions, described by
the first term in the rhs of Eq.~(\ref{kinetic0}), can be considered
independently of the other processes described by the second term
there.


\end{document}